\documentclass[showpacs,preprintnumbers,aps,prd,eqsecnum,floats]{revtex4}

\usepackage{graphicx}
\usepackage{bm}
\usepackage{epsfig}
\def\no{\nonumber}

\def\be{\begin{equation}}
\def\bea{\begin{eqnarray}}
\def\bem{\begin{multiline*}}
\def\eem{\end{multiline*}}
\def\eea{\end{eqnarray}}
\def\ee{\end{equation}}
\def\bi{\begin{itemize}}
\def\ei{\end{itemize}}
\def\cross{\times}

\def\D{\Delta}

\def\vt{\vartheta}

\def\vt{\vartheta}
\def\ltsima{$\; \buildrel < \over \sim \;$}
\def\simlt{\lower.5ex\hbox{\ltsima}}
\def\gtsima{$\; \buildrel > \over \sim \;$}
\def\simgt{\lower.5ex\hbox{\gtsima}}
\begin{document}

\title{Parameter estimation of binary compact objects with LISA: Effects
of time-delay interferometry, Doppler modulation, and frequency evolution}

\author{Aaron Rogan}
\email{roganelli@wsu.edu}
\author{Sukanta Bose}
\email{sukanta@wsu.edu}
\affiliation{%
Department of Physics \& Astronomy,
Washington State University, 1245 Webster, Pullman,
WA 99164-2814, U.S.A.
}%

\begin{abstract}

We study the limits on how accurately LISA will be able to estimate
the parameters of low-mass compact binaries, comprising white dwarfs (WDs),
neutron stars (NSs) or black holes (BHs), while battling the amplitude,
frequency, and phase modulations of their signals. These modulations will arise from LISA's 
motion relative to the source, the evolution of the source emission frequency, 
and time-delay interferometry. We show that Doppler-phase modulation aids 
sky-position resolution in every direction, improving it especially
for sources near the poles of the ecliptic coordinate system 
by a factor of at least six more than those near the ecliptic.
However, the same modulation increases the frequency estimation {\em error} 
by a factor of over 1.5 at any sky position, and at a source 
frequency of 3 mHz, which is near the high-frequency end of the ``unresolvable'' white-dwarf confusion noise. 
Since accounting for Doppler-phase modulation is absolutely essential at all LISA 
frequencies and for all chirp masses in order to avoid a fractional loss of
signal-to-noise ratio (SNR)
of more than 30\%, LISA science will be simultaneously 
aided and limited by it. 
For a source with an initial instantaneous frequency $\nu \simgt 2.5$ mHz,
searching for its chirp or frequency evolution over a one-year duration
worsens the error in the estimation of 
$\nu$ by a factor of over 3.5 relative to that of sources 
with $\nu \simlt 1$ mHz. Increasing the
integration time to 2 years reduces this relative error factor to about 2, which
still adversely affects the resolvability of the galactic binary confusion noise.
Thus, unless the mission lifetime is increased several folds, the only other
recourse available for reducing the errors is to exclude the chirp parameter 
from ones search templates. Doing so improves the SNR-normalized parameter 
estimates. This works for the lightest binaries since their SNR itself does 
not suffer from that exclusion. However, for binaries involving a neutron star, 
a black hole, or both,
the SNR and, therefore, the parameter estimation, can take a significant 
hit, thus, severely affecting the ability to resolve such members in LISA's 
confusion noise. Among the affected sources are galactic low-mass binaries 
containing a 
black hole, about which very little is known and which LISA may detect. 
Finally, we demonstrate how relative to a Michelson network a time-delay interferometric (TDI) data-combination network will have a different sensitivity sky-pattern and discuss its effect on parameter estimation.

\end{abstract}


\maketitle

\section{\label{sec:intro}INTRODUCTION}

The Laser Interferometric Space Antenna (LISA) is a proposed 
space-based broad-band gravitational wave (GW) detector funded by the National
Aeronautics and Space Administration (NASA) and the European Space 
Agency (ESA) \cite{PrePhaseA}. Unlike its extant and proposed 
earth-based counterparts, which include LIGO \cite{Abbott:2003vs}, 
VIRGO \cite{VIRGO}, TAMA \cite{TAMA}, GEO \cite{Willke:2002bs}, and LCGT \cite{Mio:2003ii},
it enjoys guaranteed sources, specifically galactic binaries, such as
AM CVn, HP Lib, WD 0957-666, RXJ1914+245, and 4U1820-30 \cite{Cutler:2002me,Nayak:2003na}. In fact, LISA will be so sensitive
that the population of galactic binaries of compact objects, which comprise
white dwarfs (WDs), neutron stars (NSs), and black holes (BHs), will create
source confusion noise that towers over LISA's instrumental noise
near the sweet spot of its observation band \cite{Bender:1997hs,hils,nelemans}. 
This can potentially affect the detection of other GW sources. 

The confusion noise from WD binaries will be especially worse below 
about 3 mHz, where a good 
fraction of it, with low signal strengths, will be unresolvable. Galactic 
binaries that are bright enough or orbiting fast enough form a potentially
resolvable population extending up to about 20 mHz. Our ability to resolve 
this latter class of sources will largely depend on how well we can estimate 
their signal
parameters. Although the development of algorithms for resolving the component
sources is still at its nascency \cite{gclean,mohanty,Crowder:2006wh}, it is already clear that the smallness of 
the parameter errors is of utmost significance to the success of any of these
algorithms. The focus of this paper is to study the role of different sources
of these errors and determine the limits on how small we can keep them.

The problem of parameter-estimation accuracy has been studied for a variety 
of LISA sources in the past. Cutler did the first systematic study 
on this subject for monochromatic sources and super-massive black hole 
mergers \cite{Cutler:98}.
Takahashi and Seto \cite{Takahashi:02} studied the problem exclusively in 
the context of low-mass compact binaries. They placed 
emphasis on the determination of the chirp parameter and its role
in unraveling the distance to a binary, following an idea due
to Schutz \cite{Schutz}.
Barack and Cutler \cite{Barack:2003fp} studied this problem in the context 
of extreme-mass-ratio inspiraling binaries. Vecchio and Wickham
\cite{Vecchio:2004ec} extended earlier parameter estimation studies 
to beyond the long-wavelength approximation by including the 
frequency-dependent modulation of LISA's beam-pattern functions arising 
at higher frequencies. They showed that invoking this approximation
results in atmost a 3\% loss in the SNR, but can have significant effects
on the parameter accuracy (anywhere from 5\% to a factor of 10) in the 
source frequency range 3mHz$\simlt \nu \simlt $10mHz.

All of these works, however, assumed that the data being filtered are
two linearly independent LISA data streams, called Michelson variables,
which measure the differential changes of length in LISA's three arms.
While these early works were instrumental in deriving useful lower-limits on the 
parameter errors of high signal-to-noise ratio (SNR) sources, 
the attainability of these limits is unclear 
because the Michelson variables will 
suffer from the presence of excess noise arising from the laser 
frequency fluctuations and optical bench motion. 
The first noise source alone is sufficient to ruin LISA's characteristic 
strain sensitivity by several orders of magnitude. Fortunately, this 
critical problem has been addressed by devising a method for combining 
LISA's data streams in software to produce data combinations that have 
these noises mitigated to the level of LISA's noise 
budget \cite{Tinto:yr,Armstrong:99,TEA}. 
This method is termed as time-delay interferometry (TDI) and will be 
introduced in Sec. \ref{sec:streams}.

In this work, like Takahashi and Seto, we consider the parameter 
estimation accuracy
of low-mass compact binaries that are not necessarily monochromatic.
But unlike that work, we base our accuracy limits on TDI variables and
do not use the long-wavelength approximation. 
We study the role of the chirp-parameter, which is tantamount
to a monotonic increase in the inspiral frequency, in affecting the 
accuracies of all signal parameters. We also quantify 
the effect of the Doppler-phase
modulation to the same end. Our study shows in what regions of the
parameter space the effects of searching for the chirp parameter or 
the effect of Doppler-phase modulation are not negligible. This can be 
useful in simplifying prototypical studies for those sources for which
these effects are small. On the other hand, parameter regions where a search 
for the chirp in a signal can hurt the accuracy of the other 
parameters estimates are identified. Moreover, we 
demonstrate that the sensitivity sky-pattern of a network of TDI variables 
is very different from that of a network of Michelson variables.
A similar difference exists in parameter accuracies achievable by these
two types of networks.
Since any realistic search will be conducted in a network of the former
type, inferences based on a network of the latter type will be off.

The layout of the paper is as follows. In Sec. \ref{sec:streams} 
we briefly introduce
a couple of TDI data combinations and show how they are obtained from
combining the frequency shifts monitored in the laser beams traversing LISA's arms.
Section \ref{sec:signal} describes the signal from low-mass compact binaries that 
are in circular orbits, but are spiraling in. The complete set of 
parameters specifying the GW signal from such sources is spelled out.
A formalism for determining the limits on parameter accuracies in
noisy measurements is given in Sec. \ref{sec:paramEst}. This introduces the 
Fisher information matrix, whose inverse gives the limits on 
the elements of the variance-covariance matrix of parameter errors \cite{Hels}. 
The effect of Doppler-phase modulation on the SNR of a signal is explained 
in Sec. \ref{sec:doppler}, and that of the source frequency evolution on the 
same quantity is explored in Sec. \ref{sec:chirp}. The parameter errors 
themselves are studied as functions of the source frequency in Sec. \ref{sec:freqDep} and as functions of sky position
in Sec. \ref{sec:errorSkyMaps}. We end with a discussion of the implications of this work
in Sec. \ref{sec:conclusion} \cite{lisaSite}.

\section{\label{sec:streams}The Data Streams}

\begin{figure}[!hbt]
\centerline{\psfig{file=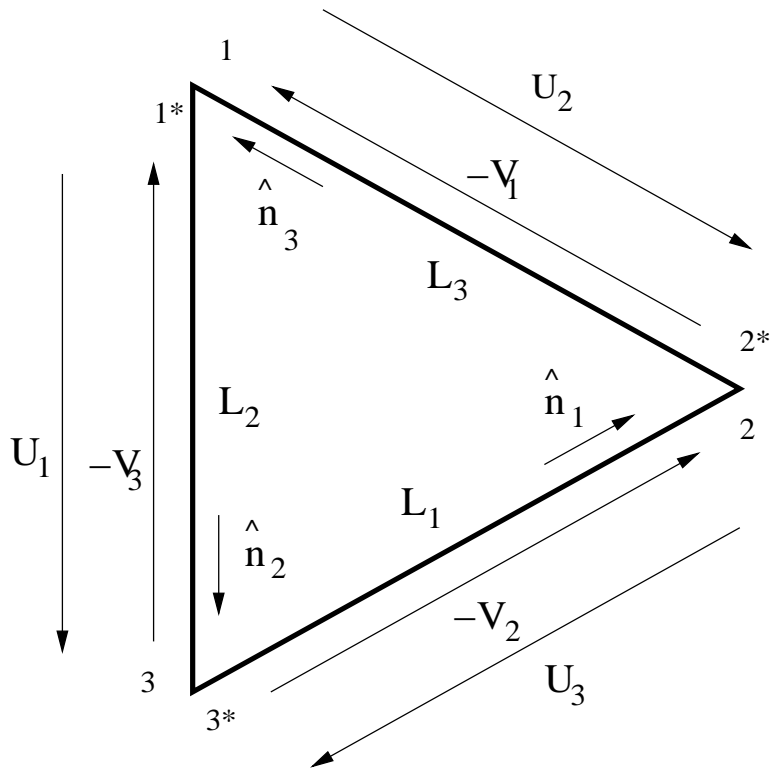,height=4.in,width=4.in}}
\caption{LISA consists of 3 spacecraft located at the vertices of
a nearly equilateral triangle. These craft exchange six elementary data
streams, labeled $U_i$ and $-V_i$. The  $U_i$ streams propagate
clockwise, whereas the $-V_i$ propagate counterclockwise.}
\label{LISAtri}
\end{figure}

The gravitational-wave information in LISA will be accessed
through a total of twelve data streams exchanged among the three spacecraft,
labeled $i=$1, 2, and 3 that are located at the vertices of an almost equilateral
triangle, as shown in Fig. \ref{LISAtri}. There, $i$ and $i^*$ denote the two
optical benches mounted in the $i$th craft. The arm-lengths of this
triangle are labeled such that $L_i$ is the length of the arm facing vertex
$i$. The unit vector, $\hat{\bf n}_i$, specifies the orientation of the $i$th
arm, and goes counterclockwise around the triangle in the figure.
Six of these beams are inter-craft beams for monitoring gravitational waves,
with 2 beams exchanged per spacecraft pair. The remaining
six beams are intra-craft beams (not shown in the figure), with 2 streams
exchanged between the two optical benches, $i$ and $i^*$, within each craft.

An impinging GW will cause shifts in the central
frequency of the laser beam exchanged between adjacent LISA craft.
However, a GW is only one of the many sources of
frequency fluctuations. There exist other noise sources that cause
appreciable shifts in the laser frequency such that GW signals would
be dwarfed in their presence. Shifts in the central frequency of
lasers can be largely mitigated in hardware in a fixed arm-length interferometer. 
However, since LISA will be unable to maintain identical arm lengths 
while the craft formation fly around the sun, the beams in each arm will
possess significant frequency fluctuation noise.
The second largest source of noise results from the mechanical vibrations 
of the optical benches on each craft.
Time-delay interferometry has been shown to reduce both laser frequency
and optical-bench motion noise to the level of LISA's noise budget, thus,
allowing LISA to obtain the sensitivity necessary to function
effectively as a gravitational wave detector \cite{Tinto:yr}.

To appreciate the basic idea behind TDI, we begin by introducing
LISA's primary data products that will contain
information about GW signals. Let $f_0$ be the central frequency of
all the lasers in LISA, then the fractional shift in laser frequency
in the beam originating at optical bench $i$ can be written as \be
C_i (t) \equiv \frac{\D f_i(t)}{f_0} \ \ . \ee 
Likewise for the
beam from the bench $i^*$. It is important to note that what is
directly measured is not the frequency fluctuation in each beam;
rather it is the fractional frequency shift in a beam from the bench
on one craft relative to that in the beam from the bench in an
adjacent craft. This is how the six inter-craft data streams,
labeled $U_i$ and $-V_i$ for ``clockwise'' and ``counterclockwise''
(see Fig. \ref{LISAtri}), are obtained. The data stream $U_i$ is
obtained by beating the fractional frequency shift in the beam from
bench $(i-1)$ relative to that in the beam originating at bench
$i^*$ \cite{indexing}. For instance, $U_1(t)
\equiv C_3(t-L_2/c)-C_1(t)$. The remaining two streams, namely,
$U_2$ and $U_3$, can be obtained by cyclic permutation of the
indices in the $U_1$ expression. In the rest of the paper, we set
the speed of light, $c$, to unity. The $-V_i$ data streams are
obtained in a similar fashion by measuring the fractional frequency
shift in the beam from bench $(i+1)^*$ relative to that originating
at bench $i$. The time-shift operator $\zeta_i$ is defined by its
action on a data stream $x(t)$ as \cite{Rogan:2004wq}: \be
\label{operator} \zeta_i x(t)= x(t-L_i) \ \ , \ee where the label
$i$ denotes the arm along which the time-shift is affected. Then,
$U_1(t)$ can be written alternatively as $U_1(t)=\zeta_2
C_3(t)-C_1(t)$ and, similarly, for the other streams.

The intra-craft beams exchanged between adjacent optical benches on
every craft provide additional information about optical-bench motion
noise that is common to multiple data streams. Although there are
a total of 6 intra-craft beams, the two beams per craft can be
combined in such a manner so as to produce a single data stream
containing all the relevant information regarding these noise
sources. This can be seen easily by examining a single craft's
internal data streams. At craft 1, by beating the fractional
frequency shift in the beam from bench 1 relative to that in the
beam from bench $1^*$, one forms the data stream,
\be\label{intracraft} W_1 = \left(C_1-\hat{\bf n}_3\cdot {\bf
v}_1\right) - \left(C_{1^*}+\hat{\bf n}_2\cdot {\bf v}_{1^*}\right)
+ \hat{\bf n}_3\cdot {\bf u}_1 + \hat{\bf n}_2\cdot {\bf u}_{1^*}
\,. \ee where ${\bf u}_{i,i^*}$ are the random velocities of the
proof masses on benches $i$ and $i^*$, respectively, and ${\bf
v}_{i,i^*}$ are the random velocities of the optical benches
themselves. Two other intra-craft data combinations, $W_2$
and $W_3$, can be obtained by the cyclic permutation of indices in
the above expression.

There exists a set of polynomial operators, $p^A_i$, $q^A_i$, and
$r^A_i$, that when acting on the nine basic data streams (viz., six
inter-craft streams, $U_i$, $V_i$, and three intra-craft streams,
$W_i$) allow one to construct several time-delayed data combinations
that are  free from the dominating noise of the laser-frequency
fluctuation and optical bench motion. This technique first developed by
Tinto and Armstrong \cite{Tinto:yr} is referred to
as time-delay interferometry (TDI) and the resulting data streams are known
as TDI variables. Subsequently, it was shown by Dhurandhar et
al. \cite{Dhurandhar:2001kx} that a host of such laser-frequency
canceling streams exist and form the module of syzygies. In an
earlier work \cite{Rogan:2004wq}, it was shown that any of these
data combinations, or pseudo-detectors, obtained from operating with
these polynomials on the basic streams can be expressed as:
\be
\label{detector} x^{A} = {\rm Trace}\left[{\bf e}^A\cdot {\bf
Z}\right] \ \ , \ee where $A$  is the pseudo-detector index and
\be
{\small {\bf e}^A \equiv \left(\begin{array}{ccc}
p^A_1 & p^A_2 & p^A_3 \\
q^A_1 & q^A_2 & q^A_3 \\
r^A_1 & r^A_2 & r^A_3
\end{array}\right) }
\quad{\rm and}\quad {\small {\bf Z} \equiv \left(\begin{array}{ccc}
V_1 & U_1  & W_1 \\
V_2 & U_2  & W_2 \\
V_3 & U_3  & W_3 \\
\end{array}\right) \,.}
\ee
Although there are many sets of noise-canceling pseudo-detectors, one
well-studied set of pseudo-detectors is
defined by the choice \cite{Prince:2002hp,Nayak:2003na,Rogan:2004wq}:
\be
{\small {\bf e}^1 = \left(\begin{array}{ccc}
1 & \zeta_3 & \zeta_2\zeta_3 \\
1 & \zeta_1\zeta_2 & \zeta_2 \\
1+\zeta_1\zeta_2\zeta_3 & \zeta_3+\zeta_1\zeta_2 &  \zeta_2+\zeta_1\zeta_1\\
\end{array}\right) \>,}
\quad {\small {\bf e}^2 =\left(\begin{array}{ccc}
\zeta_1\zeta_2 & 1 & \zeta_1 \\
\zeta_3 & 1 & \zeta_2\zeta_3 \\
\zeta_3+\zeta_1\zeta_3 & 1+\zeta_1\zeta_2\zeta_3 & \zeta_1+\zeta_2\zeta_3 \\
\end{array}\right) \>,}
\ee and \be {\small {\bf e}^3 = \left(\begin{array}{ccc}
\zeta_2 & \zeta_2\zeta_3 & 1 \\
\zeta_1\zeta_3 & \zeta_1 & 1 \\
\zeta_2+\zeta_1\zeta_3 & \zeta_1+\zeta_2\zeta_3 & 1+\zeta_1\zeta_2\zeta_3 \\
\end{array}\right) \,.}
\ee
These pseudo-detectors were termed as $\alpha$, $\beta$, and $\gamma$,
respectively, by Armstrong et al. \cite{Armstrong:99,DhurandharCombo}.

It is important to note that the data combinations above are not
statistically independent. However, it is possible to create noise-independent
pseudo-detectors by diagonalizing the noise-covariance matrix of the above
combinations. One such triplet is \cite{Prince:2002hp}:
\be\label{AET}
A \equiv {1\over \sqrt{2}}(-\alpha+\gamma)\\
\quad E \equiv {1\over \sqrt{6}}(\alpha -2\beta + \gamma)\\
\quad T \equiv {1\over \sqrt{3}}(\alpha +\beta +\gamma)
\ee
where $T$ is akin to a Sagnac response, which is devoid of a
GW signal. The data combination $A$ is not to be confused with the
pseudo-detector {\em index} $A$: Unlike the former, the latter will
appear only as a superscript or subscript of another symbol.
What is most important to note is that the eigenvalues of the
above-mentioned noise-covariance matrix are degenerate, with $A$ and $E$
both having the same noise variance. Therefore, any pair of pseudo-detectors
that lie in the plane normal to $T$ and are orthogonal to each other will be
noise-independent as well. An infinity of such pairs are obtainable by simple
rotations about $T$. Unless specified otherwise, for the remainder of this paper, the pseudo-detector triplet or {\em network} considered is 
\be\label{AETbar} 
\bar{A}=\cos(\pi/3)A-\sin(\pi/3)E\equiv x^1 \quad \bar{E}=-\sin(\pi/3)A-\cos(\pi/3)E\equiv x^2 \quad \bar{T}=T\equiv x^3 \,. \ee 

An alternative data combination that has been studied in the literature is the pair
of Michelson variables
\bea \label{Michelson}
h^I(t)&=&(\varepsilon_1(t)-\varepsilon_2(t))  \ \ , \nonumber \\
h^{II}(t)&=&(\varepsilon_1(t)+\varepsilon_2(t)-2\varepsilon_3(t)) \ \ ,
\eea
where $\varepsilon_i$ is the strain in the $i$th arm. These variables suffer
from the frequency fluctuation and optical bench motion noise and
will not be useful in actually detecting signals once LISA is operational, except
at very low frequencies. However, owing to their simple relation
to GW polarization components \cite{Cutler:98}, they have been found to be a 
useful starting point for data analysis studies.

As discussed in Ref. \cite{Rogan:2004wq}, the three noise-independent first
generation TDI variables can be expressed as
\be \label{strainInData}
x^{A}(t) = n^A(t) + h^A(t) \ \ ,
\ee
where
\be\label{strainTime}
h^A(t) =  \sum_{i=1}^{3}\left[p^A_iV_i^{\rm GW}(t)
+q^A_iU_i^{\rm GW}(t)\right]
\ee
is the gravitational-wave signal in the $A$th
pseudo-detector and $n^A(t)$ is the remaining instrumental 
noise in it. (Note that the sum on the 
right-hand side adds up to zero for $A=3$.) The noise is assumed to have a
Gaussian probability distribution with a zero mean. The 
noise covariance matrix elements are given as follows:
\be\label{noise}
\overline{\tilde{n}^{A*}(f)\tilde{n}^{B}(f')}
=\frac{1}{2}P^{(A)}(f)\delta(f-f')\delta^{AB} \ \ ,
\ee
where $P^{(A)}(f)$ is the one-sided noise power-spectral density (PSD)
of the $A$th pseudo-detector.

The noise PSDs for the above mentioned pseudo-detectors can be expressed as
\cite{Prince:2002hp,RoganCorr1}:
\bea
\label{noisePSD} P^{(1)}(f) &=& P^{(2)}(f) = 8\sin^2 (\pi f
L)\big\{2\left[3+2\cos(2\pi f L)+\cos(4 \pi f
L)\right] P^{\rm{proof}}
\no\\
&&\quad \quad \quad\quad
+[2 + \cos(2\pi f L)]P^{\rm{shot}}\big\}\ \ ,\no\\
P^{(3)}(f) &=& (2+4\cos^2(2 \pi f L))\left[(4\sin^2(\pi f L)P^{\rm{proof}}
+ P^{\rm{shot}}\right]\,, \eea
where the noise PSDs arising from the photon-shot noise and the proof-mass
noise are estimated to be $P^{\rm{shot}}=1.8 \cross 10^{-37}[{f/1 \rm{Hz}}]^2
{\rm Hz}^{-1}$ and $P^{\rm{proof}}=2.5 \cross 10^{-48}[{f/1\rm{Hz}}]^{-2}
{\rm Hz}^{-1}$, respectively \cite{PrePhaseA}.
The proof-mass noise affects LISA's sensitivity at
low frequencies, while the shot noise affects it at higher
frequencies. Note that
$P^{(1)}(f)$ is the noise PSD for pseudo-detectors $\bar{A}$ as well
$A$; similarly, $P^{(2)}(f)$ is the noise PSD for pseudo-detectors
$\bar{E}$ as well $E$, and so on.

\begin{figure}[!hbt]
\centerline{\psfig{file=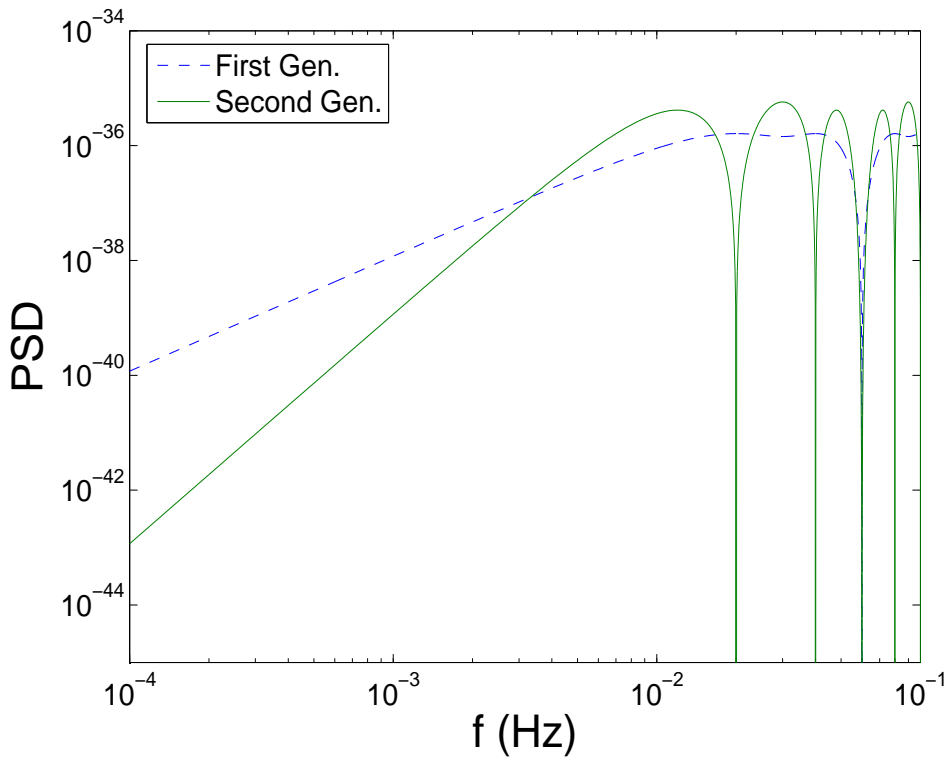,height=4.in,width=5.in}}
\caption{Plots of the first and second generation noise
power-spectral densities (in units of Hz$^{-1}$).} \label{fig:noisePSD}
\end{figure}

The TDIs introduced above are the so-called first
generation TDIs. These TDIs assume that the light travel time for
data stream $V_i$ and data stream $U_{(i+1)}$ are identical.
Realistically, however,
the rotational and orbital motion of LISA would prevent the noise contribution of the
laser-frequency fluctuations from being mitigated to the level of the
secondary noises \cite{Cornish:2003tz,Shaddock:2003dj,Tinto:2003vj}.
In order to tackle this problem, new
pseudo-detectors were introduced as simple differences of their first
generation counterparts, appropriately time-shifted:
\cite{Tinto:2003gwdaw}:
\be \label{secondGen}
h^A_2(t)=h^A(t)-h^A(t-3L) \ \ ,
\ee
where the subscript $2$ indicates that these are {\em second generation}
data combinations. One can also define the second generation noise PSD as:
\be \label{secondGenNoise}
P^{(1),(2),(3)}_2(f)=4\sin^2(2\pi f L)P^{(1),(2),(3)}(f).
\ee
Again, the subscript $2$ denotes that these are second generation
noise PSD's. The first and second generation noise PSDs are compared in
Fig. \ref{fig:noisePSD}. 

In the rest of the paper, we will discuss parameter estimation
of binary inspiral signals based on the first generation data sets
(unless otherwise specified). The above transformation can be
implemented straightforwardly to get the estimates for the second
generation data sets.

\section{\label{sec:signal}The Gravitational-wave Signal}

Consider a GW source located at a distance $R$ relative to the solar system
barycenter (SSB). Let $(\theta,\phi)$ denote the sky position 
of the source in the ecliptic coordinate system
centered at the SSB. Then the sky-position vector is given by
$\hat{\bf w} = (\sin \theta \cos \phi, \sin \theta \sin \phi, \cos\theta)$.
In the transverse-traceless gauge \cite{MTW}, the space-space part of
the metric perturbation due to a gravitational wave emitted by this source
and observed at the spacetime location (t, ${\bf r}$) can be expressed as
\be \label{perturbation}
h_{mn}(t, {\bf r}) = h_+(t-\hat{\bf w} \cdot
{\bf r})(\theta_m \theta_n-\phi_m \phi_n) + h_\cross(t-\hat{\bf w}
\cdot {\bf r})(\theta_m \phi_n+\theta_n \phi_m) \ \ ,\ee
where $m$ and $n$ are spatial indices, $\theta_m$ and $\phi_n$ are components
of unit-normal vectors $\hat{\mbox{\boldmath $\theta$}}$ and $\hat{\mbox{\boldmath $\phi$}}$, respectively, that define a transverse plane with respect to
the wave-propagation vector, $-\hat{\bf w}$, and $h_+$ and $h_\cross$
are the two polarization components of the impinging wave.
Therefore, the strain in LISA's $i$th arm is characterized by the contraction
of the arm's ``detector'' tensor, $n_i^m n_i^n$, with the perturbation
matrix $h_{mn}(t, {\bf r})$:
\be \label{ithStrain}
h_i(t) = h_+(t) \xi_{i +}(\theta, \phi) +
h_\cross(t) \xi_{i \cross}(\theta, \phi) \ \ ,\ee
where
\be \label{Xis}
\xi_{i +}(t)=(\hat{\mbox{\boldmath $\theta$}}
\cdot \hat{\bf n}_i(t))^2-(\hat{\mbox{\boldmath $\phi$}}  \cdot \hat{\bf n}_i(t))^2 \hspace{10pt} \quad {\rm and} \quad
\xi_{i \cross}(t)=2(\hat{\mbox{\boldmath $\theta$}} \cdot \hat{\bf n}_i(t))(\hat{\mbox{\boldmath $\phi$}} \cdot \hat{\bf n}_i(t))
\ \ , \ee
define that arm's beam pattern.

In this paper, we consider how well one would be able to estimate the
parameters of a nearly monochromatic source, such as a low-mass compact
binary, in the LISA data. Let the masses of the two stellar members of
these sources be $m_1$ and $m_2$. Then, in the quadrupole approximation, the
amplitude of its signal at the SSB can be expressed as
\cite{Rogan:2004wq}
\be \label{amplitude}
H(\omega) = 1.188\cross 10^{-21} \left[{\mathcal{M}\over
M_\odot} \right]^{5/3} \left[{R\over 1{\rm kpc}}\right]^{-1}
\left[{\omega \over 2 \pi \hspace{2pt} {\rm mHz}}\right]^{2/3}.
\ee
where $\mathcal{M} = m_1^{3/5} m_2^{3/ 5}/ (m_1+m_2)^{1/5}$ is
its chirp mass, $R$ is its luminosity distance, and $\omega \equiv 2\pi \nu$
is its angular frequency.

For a sufficiently high-mass binary, the inspiral of its members will
cause the source frequency to increase perceptibly in the LISA band. On the other
hand, for small chirp mass and emission frequency, as studied here, the rate of
frequency evolution is approximately \cite{Rogan:2004wq}
\be \label{evolution}
\dot{\omega} = {48\over5}\left({G \mathcal{M} \over 2}\right)^{5/3}
\omega^{11/3}, \ee
where $G$ is the universal gravitational constant.

Both the amplitude and the phase of a GW signal
will undergo modulation due to LISA's changing orientation and motion
relative to the source. The amplitude modulation is determined by
the sky position angles and the source orientation. The latter can be specified
in terms of the polarization angle, $\psi$, and the inclination,
$\iota$, of the binary's orbit relative to the line of sight. As was shown
in Ref. \cite{Rogan:2004wq}, this modulation in the $A$th pseudo-detector,
arising from LISA's $i$th arm, is captured by the complex {\em extended}
beam pattern functions,
\be \label{beampattern}
E^{A}_{i}:= T^{~\rho}_2 \hspace{3pt} D^{~A}_{\rho \hspace{1pt} i} \ \ ,
\ee
where the sum over the values of $\rho=\pm 2$ is implicit. The advantage of
the above decomposition of the extended beam patterns is that all the
source-orientation dependence is limited to the
Gel'Fand functions \cite{GMS,Bose:1999pj,Pai:2000zt},
\be T_2^{~\pm
2}\left(\psi,\iota,0\right) = \frac{1}{4}\left(1\pm
\cos\iota\right)^2 \exp\left(\mp {\rm i}2\psi\right) \,.\ee
The influence of the source location and the arm orientation on the
amplitude modulation is factored in
\be \label{Ds} D_{\pm 2 \hspace{1pt} i}^A =
{\omega L_i \over 2} |M_i^A|(\xi_{i+}\mp i \xi_{i
\cross}) \ \ ,
\ee 
where 
\be \label{MiA} M_i^A=q^A_{i-1}{\rm sinc} \left
(\omega L_{i} {(1+ \hat{\bf w} \cdot \hat{\bf n}_i)\over 2}
\right) -p^A_{i+1}{\rm sinc} \left (\omega L_i{(1- \hat{\bf w}
\cdot \hat{\bf n}_i)\over 2} \right) \,.
\ee
As was shown in Ref. 
\cite{Rogan:2004wq}
the strain in the $A$th
pseudo-detector can be expressed as:
\be \label{GWsignal} h^A (t) = H(\omega)
\sum_{i=1}^3 \Re \left[ e^{-{\rm i} \delta} E_i^{A*}(t) S_i^A(t)\right]
\ \ ,\ee
where $\delta$ is the initial phase and
\be \label{signal}
S_i^A \equiv e^{{\rm i}\omega (t-L_i\tau_i)+{\rm i}({1\over2}\dot{\omega}t^2+\sigma_i^A)+{\rm i}\phi_D(t)}
\ee
describes the phase evolution of the signal. Above, $\sigma_i^A = \arg (M^A_{i})$
denotes the modulation in the phase arising from LISA's changing beam patterns, and
\be \label{timedelay}
\tau_i \equiv{1 \over 2} \left(1-{\hat{\bf w}
\cdot \hat{\bf r}_i \over \sqrt{3} }\right ) \ \ , \ee
where $L_i \hat{\bf w} \cdot \hat{\bf r}_i/\sqrt{3}$ is the delay in the
arrival of the signal at the $i$th craft relative to LISA's centroid. Finally,
\be \label{doppler}
\phi_D(t) = \omega d_\odot \sin\theta\cos\left(\frac{2\pi t}{P}-\phi\right)
\ee
gives rise to the Doppler-phase modulation owing to LISA's motion relative to the source.
Here, $d_\odot$ (= 1 AU) is the average distance to the sun from LISA's centroid and
$P$ (=1 year) is the time taken for LISA to complete one full orbit around the sun.

The strain expression given in Eq. (\ref{GWsignal}) will be used below
to ascertain how well we can estimate a slightly chirping ($\dot{\omega}\neq 0$)
binary's parameters with LISA.
The above discussion shows that such as signal will be
characterized by eight parameters,
$\mbox{\boldmath $\vt$}=
\{ T^2\dot{\nu}, H,\delta, \psi, \cos\iota, \phi, \cos\theta, T\nu\}$,
such that, $\vt^0 = T^2\dot{\nu}$, $\vt^1 = H$, ..., and $\vt^7=T{\nu}$.
Note that all parameters are dimensionless.

\section{\label{sec:paramEst}Estimating signal parameters}

In estimating the parameters of a signal, an observer has to confront the
inherently noisy nature of the detector or receiver, which introduces a degree
of randomness in the parameter measurements. Thus, an ensemble of
multiple copies of the detector will return a distribution of parameter
values that is affected by the characteristics of the detector noise.
The spread in the distribution indicates the noise-limited accuracy of
the detector.

Before we define our estimator and determine its accuracy,
note that even in the absence of instrumental noise, a typical detector
has physical limits on its measurement accuracy. In this context,
it is important to note that there are two important time-scales
that affect LISA's observations: The first arises from LISA's orbital 
baseline, the mean light-travel time for
which is close to 1000s. Thus, the frequency of a wave with wavelength 
equal to this distance is about 1mHz.
The other scale arises from LISA's 5 million-km arm-length, which
corresponds to a light-travel time of 16.67s. Thus, waves with frequencies
that are odd-integral multiples of 0.03Hz will drive the two craft at the
ends of an arm out of phase by 180 degrees. This makes LISA's sensitivity
oscillatory at the higher end of its band (see Eqs. (\ref{noisePSD}) and
Fig. \ref{fig:noisePSD}.)

To ascertain how large the noise-limited errors will be in the parameter 
values, we take the parameter values themselves to be the maximum likelihood 
estimates (MLEs). Owing to noise, the MLEs,
$\hat{\mbox{\boldmath $\vt$}}$, can deviate from the true values
of the signal parameters and can influence
the determination of other parameters.
The magnitude of these deviations and influences can be quantified by
the elements of the
variance-covariance matrix, $\gamma^{mn}=\{\hat{\vt}^m,\hat{\vt}^n\}$ \cite{Hels}.

A relation between the $\gamma^{mn}$ and the signal is available through
the Cramer-Rao inequality, which dictates that
\be
\parallel {\mbox{\boldmath $\gamma$}}\parallel ~\geq~ \parallel {
\mbox{\boldmath $\Gamma$}}\parallel^{-1} \ \ ,
\ee
where $\mbox{\boldmath $\Gamma$}$ is the Fisher information matrix:
\bea
\label{Fisher} \Gamma_{mn} &=& \sum_{A=1}^2
\bigg \langle \partial_m h^A(\mbox{\boldmath $\vt$}),
\partial_n h^A(\mbox{\boldmath $\vt$}) \bigg \rangle_{(A)} \no \\
&\equiv& \sum_{A=1}^2 {2\over P^A(f)} \int_{0}^{T}
\left[\partial_m h^A(t; \mbox{\boldmath $\vt$})\right]
\left[\partial_n h^A(t; \mbox{\boldmath $\vt$})\right] dt
 \,.
\eea
In this paper, we determine the values of $\Gamma_{mn}$
for signals from low-mass compact binaries in LISA data. Therefore,
$\Gamma_{mm}^{-1/2}$ gives the lower bound on $\Delta\vt^m$
(which is the expected random error in the MLE of $\vt^m$).
The two are equal in the limit of large SNR.

Since the amplitude and the initial phase are extrinsic parameters, 
one can analytically maximize the likelihood ratio with respect to 
them \cite{Rogan:2004wq}. Here, we assume that this has been done. In 
such a case, the Fisher information matrix is six-dimensional. The 
errors in the phase and the amplitude can be derived in terms of the
errors in the remaining six parameters as demonstrated, e.g., in 
Ref. \cite{Jaranowski:2005hz}. This is the method we use to compute the error
in $H$. One can similarly infer the error in $\delta$.

The errors in the sky-position angles will be presented in terms of the
error in the measurement of the sky-position {\em solid angle}, defined as:
\be \label{angularRes}
\Delta\Omega_S = 2 \pi \left(\Delta\vt^5\Delta\vt^6 - 
\langle\Delta\vt^5\Delta\vt^6\rangle\right) \,.
\ee
We define the error in the measurement of the source-orientation
solid angle similarly:
\be \label{orientationRes}
\Delta\Omega_L = 2 \pi \left(\Delta\vt^3\Delta\vt^4 - 
\langle\Delta\vt^3\Delta\vt^4\rangle\right) \ \ ,
\ee
where the subscript $L$ denotes the orbital angular momentum of the binary.

\subsection{\label{subsec:paramNorm}Scaling of parameter errors with SNR}

Although our expressions above can be used to compute the error estimates
for any parameter values, we illustrate our results for specific values of
some of these parameters: Unless specified otherwise, we use
an integration time of one year, an emission frequency of $\nu =3$ mHz,
and source orientation angles $\psi = {\pi /3}$ and $\iota = {\pi / 4}$.
The emission frequency is chosen to be large enough so that
the Doppler-phase modulations have a significant effect.
The source orientation is chosen somewhat arbitrarily, except that
we ensure that the signal is not linearly polarized.

All the parameter estimates studied below are obtained by normalizing the estimates
given in Eq. (\ref{Fisher}) by the network sensitivity,
$\left[\sum_{A=1}^2\langle h^A,~h^A\rangle_{(A)}\right]^{-1}$.
Thus, the parameter-estimate values shown in the plots here are for a signal
with an SNR of 1. To get an estimate for an arbitrary SNR, one simply multiplies
the value shown in the plots by the SNR scale given in Table \ref{tab:snrScales}.

\begin{table}[!hbt]
\caption{The table lists how the errors in different parameters scale with the
SNR. Since all our error plots are given for an SNR of unity, to assess the
error for any other SNR one simply needs to multiply the plot value with
the scale read from this table.
}
\vskip 5pt
\begin{tabular}{|c|c|c|c|c|c|}
\hline &   $\quad\Delta \Omega_S\quad$   &  $\quad\Delta
\Omega_L\quad$   &  $\quad\Delta H/H\quad$ &
$\quad T \Delta \nu \quad$  & $\quad T^2 \Delta \dot \nu \quad$   \\
\hline
SNR scale  &  $\quad{\rm SNR}^{-2}\quad$  &  $\quad{\rm SNR}^{-2}\quad$
& $\quad{\rm SNR}^{-1}\quad$& $\quad{\rm SNR}^{-1}\quad$& $\quad{\rm SNR}^{-1}\quad$ \\ \hline
\end{tabular}
\label{tab:snrScales}
\end{table}

An alternative way of understanding the parameter estimate normalization is
through the Fisher information matrix.
The idea here is to first divide the Fisher information matrix by the
network template norm, $\sum_{A=1}^2\langle h^A,~h^A\rangle_{(A)}$, as follows:
\be \label{FisherNorm1}
\widehat{\Gamma}_{mn}=\frac{\Gamma_{mn}}{{\sum_{A=1}^2\langle h^A,~h^A\rangle_{(A)}}}
\,.\ee
One then obtains the normalized error estimates from the covariance matrix
derived by inverting $\widehat{\Gamma}_{mn}$.

\section{\label{sec:doppler}Doppler-phase modulation}

\begin{figure}[!hbt]
\centerline{\psfig{file=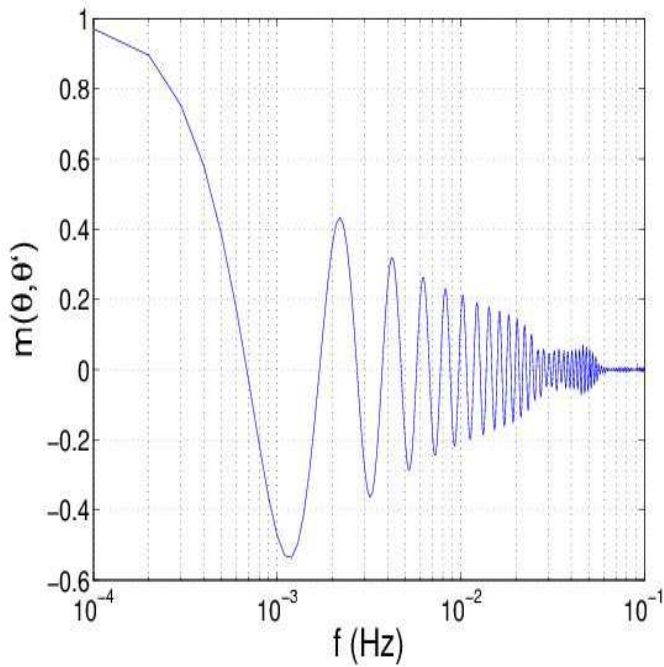,height=3.5in,width=3.5in}}
\caption{This is a plot of the fitting factor between two similar
templates, one with the Doppler-phase modulation included and
the other without it.} \label{dopplermismatch}
\end{figure}

When a template does not track the Doppler-phase modulation of a signal,
over time it gets increasingly phase-incoherent relative to the latter.
We quantify the degree of coherence between the two by the fitting factor:
\be \label{mismatchEq}
m(\mbox{\boldmath $\vt$},\mbox{\boldmath $\vt$}') = {\sum_{A=1}^2\langle h^A(\mbox{\boldmath $\vt$}), h^A(\mbox{\boldmath $\vt$}')
\rangle_{(A)} \over \sqrt{\sum_{B=1}^2\langle h^B(\mbox{\boldmath $\vt$}), h^B(\mbox{\boldmath $\vt$}) \rangle_{(B)} \>\sum_{C=1}^2\langle
h^C(\mbox{\boldmath $\vt$}'), h^C(\mbox{\boldmath $\vt$}') \rangle_{(C)}}}
\ \ ,\ee
where the value of the ``template parameter'' $\mbox{\boldmath $\vt$}'$ is, in
general, different from that of the ``signal parameter''
$\mbox{\boldmath $\vt$}$.
The plot of the above function, with $h^A(\mbox{\boldmath $\vt$}')$
as the template without Doppler shifting and $h^A(\mbox{\boldmath $\vt$})$ as the
signal with that shifting accounted for, is given in
Fig. \ref{dopplermismatch}.
As can be inferred from that figure, when the Doppler-phase modulation is 
not accounted for in the search templates, the SNR suffers an appreciable drop
across the entire LISA band. It also exhibits large oscillations with a period
that depends on LISA's orbital speed and that decreases linearly with increasing
source frequency, as is characteristic of Doppler frequency shifting.
It is important to note the locations of the
nodes in the plot: The first minimum is at approximately $f=$1/2L, which occurs
when the phases of a gravitational wave impinging along an arm are different
by 180$^\circ$ at the two craft at the two ends of that arm.

In conclusion,
since the effect of discounting Doppler-phase modulation is so marked on the SNR at any LISA frequency of interest,
one must account for its effect on parameter estimation throughout the LISA band.

\section{\label{sec:chirp}Effect of source frequency evolution}

{}For binary compact object that are spiraling in fast enough, LISA
will be able to measure their chirp mass and luminosity distance.
Expression (\ref{amplitude}) shows that the amplitude, $H(\omega)$, of a signal
from such an object depends jointly on $r$ and ${\cal M}_c$.
However, neither of these parameters
affects any other part of a {\em monochromatic} (or $\dot{\nu}$)
signal, as given in Eq. (\ref{GWsignal}). Thus, one can not
separately estimate either of them purely from the amplitude.
Nevertheless, if a signal has an appreciable amount of inspiral or frequency evolution, defined in
Eq. (\ref{evolution}), then the measurement of $\dot{\nu}$, along with
$H(\omega)$, determines both the chirp mass and the source distance.

In order to ascertain the individual masses of the binary,
additional information is necessary.
Traditional astronomical methods using
optical or radio measurements can be used to help identify the total
mass of the binary system. However for many unknown sources or
optically invisible sources this is not an option. In some cases
where the binary orbit is noticeably eccentric, the accurate tracking of
the emission frequency can provide information on the total mass of the
system. This datum, together with the chirp mass, can be used to infer
the individual masses of the system \cite{Seto:2001pg}.

\begin{figure}[hbt]
\centerline{\psfig{file=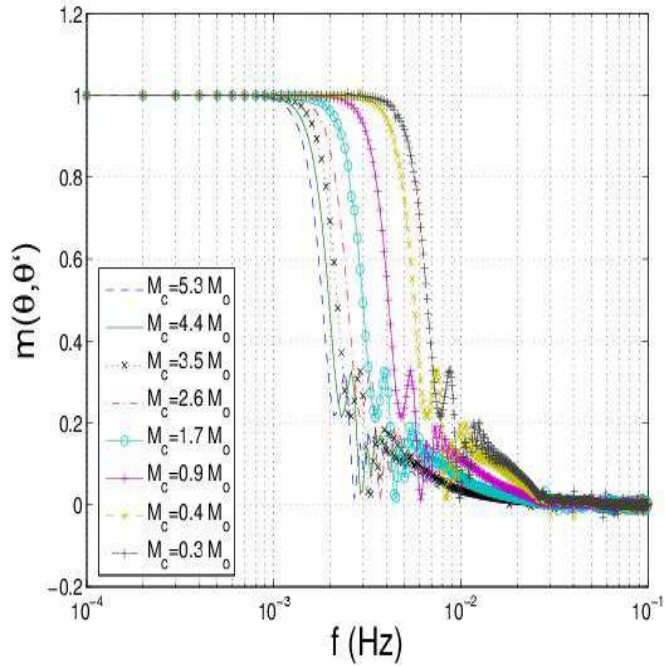,height=3.5in,width=3.5in}}
\caption{Fitting factors of low-mass compact binaries with a range of chirp masses are plotted as functions of frequency. Here the fitting factor is between two template families, one with a physical set of
eight parameter values and another with the chirp ($\vt^0$) set to zero, 
artificially, for all source frequencies.
For the above plots, the sources are located on the ecliptic, with $\theta=\pi/2$, $\phi=\pi/2$, and their orientation is $\iota=\pi/4$,
$\psi=\pi/3$.
As a reference, an equal-mass binary, with $m_1=m_2\equiv 0.35~M_\odot$ has ${\cal M}_c = 0.3~M_\odot$. Fitting factors for equal-mass binaries with $m=$1,2,3,...6$~M_\odot$ can also be read from here.
}
\label{omegadotmismatch}
\end{figure}

Before analyzing the quantitative effects of a chirping waveform on
parameter estimation
we briefly study the role that a non-negligible $\dot{\nu}$ plays in searching for a signal.
Specifically, we use the fitting factor defined in Eq. (\ref{mismatchEq}) to
determine the fractional loss in the SNR while searching for
a chirping source with monochromatic (i.e., $\dot{\nu}=0$) templates.
When subtracted from unity, the above function measures the fractional
drop in the SNR owing to a parameter mismatch of $\D\mbox{\boldmath $\vt$}
\equiv \mbox{\boldmath $\vt$} - \mbox{\boldmath $\vt$}'$.
In Fig. \ref{omegadotmismatch}, we plot
$m(\mbox{\boldmath $\vt$},\mbox{\boldmath $\vt$}')$ with
$\D\vt^m = 0$, for all $m \neq 0$; the ``template parameter'', $\vt'^{\>0}$,
is set to zero, while the ``signal parameter'', $\vt^0$, is determined by
Eq. (\ref{evolution}) for any given emission frequency and chirp mass. 
For any given chirp mass, note how rapidly
the SNR drops as a function of frequency.
Also, since at any given frequency the chirping causes a smaller mismatch between
the two template families for a smaller chirp mass, the SNR drop occurs at a higher frequency for a smaller chirp mass.
Just like Fig. \ref{dopplermismatch}, this fitting factor too has
oscillatory features.
However, owing to the polynomial frequency behavior of $\dot{\nu}$, the
oscillations here do not exhibit the linear decrease in the node-spacing
found in Fig. \ref{dopplermismatch}.

We now turn our attention to the role of $\dot{\nu}$ in the estimation of
the signal parameters. As shown in Sec. \ref{sec:paramEst}, the expansion of the
template parameter space in order 
to include another parameter affects the parameter
estimation through the Fisher information matrix, which is a different
construct than the fitting factor.
Once a detection has been made, the error values presented
here provide the standard deviation that should be expected for an 
ensemble of signal measurements made by LISA.
However, one still needs to establish for what
ranges of $\omega$ and ${\cal M}_c$ it is possible to measure
a non-vanishing chirp in a signal. Outside these ranges, 
including $\dot{\omega}$ in ones template-parameter space
can result in worse estimates for other parameters. A smaller parameter
volume for estimating parameters is also desirable because it leads to a
lesser computational burden: This is more so
since the problem of detection in the presence of LISA's source confusion
noise is intricately linked with the problem of parameter estimation.

\begin{figure}[!hbt]
\centerline{\psfig{file=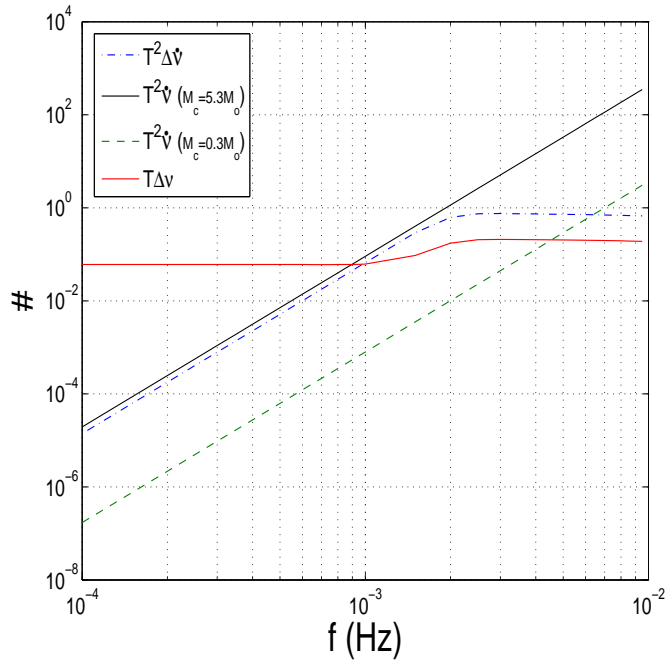,height=3.5in,width=3.5in}}
\caption{These plots illustrate the values (in number of cycles) 
of the chirp parameter for two values of the chirp mass 
(${\cal M}_c = 0.3M_\odot$ and $5.3M_\odot$), 
along with the errors in determining the
chirp and frequency for an SNR of 10. All plots are for an observation period of $T=1$ year. Note how for the larger chirp mass the error in determining the chirp remains comparable to the value of the chirp itself up to about 1.5mHz, before improving at higher frequencies.
}
\label{freqplots3}
\end{figure}

\begin{figure}[!hbt]
\centerline{\psfig{file=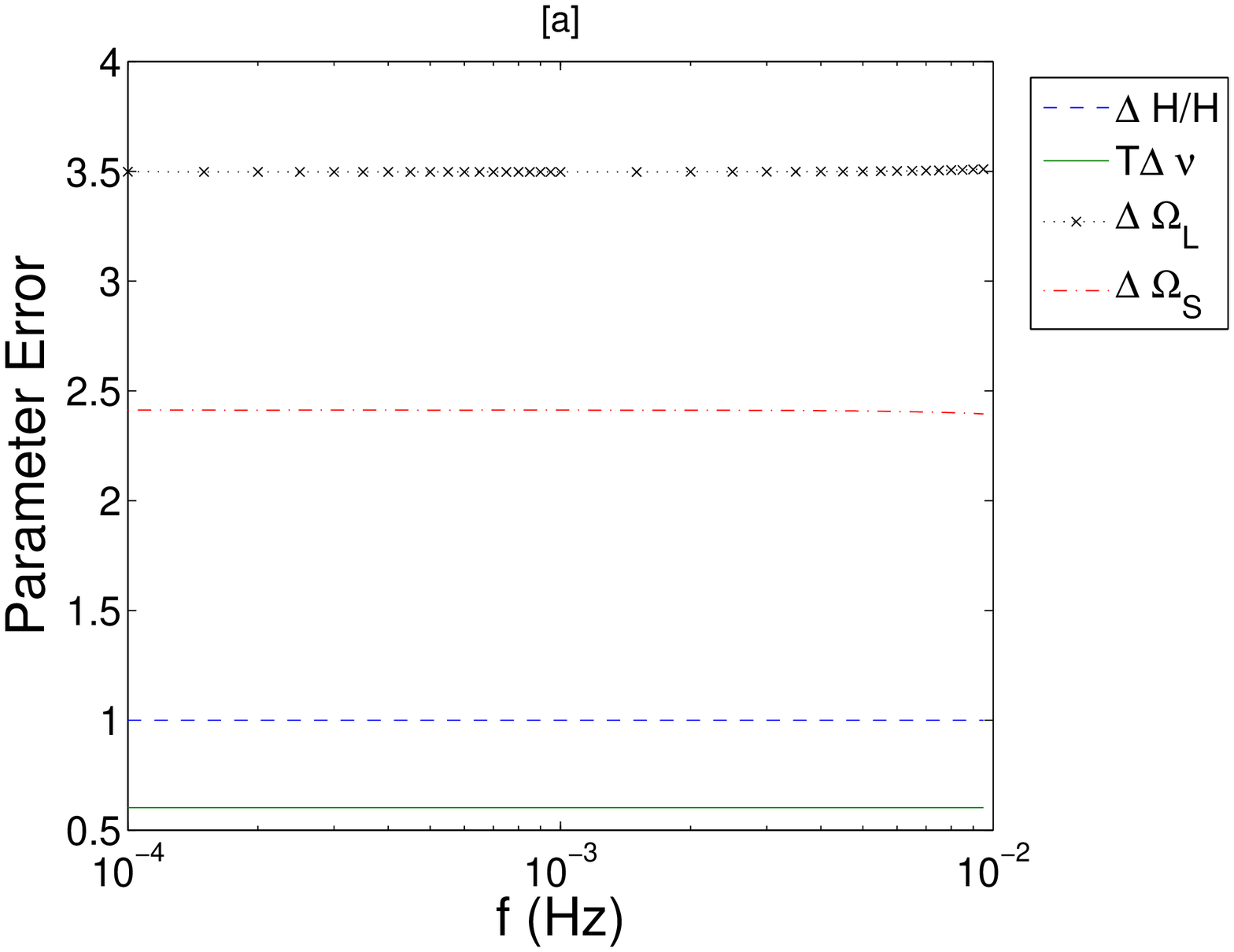,height=3.in,width=3.in}\psfig{file=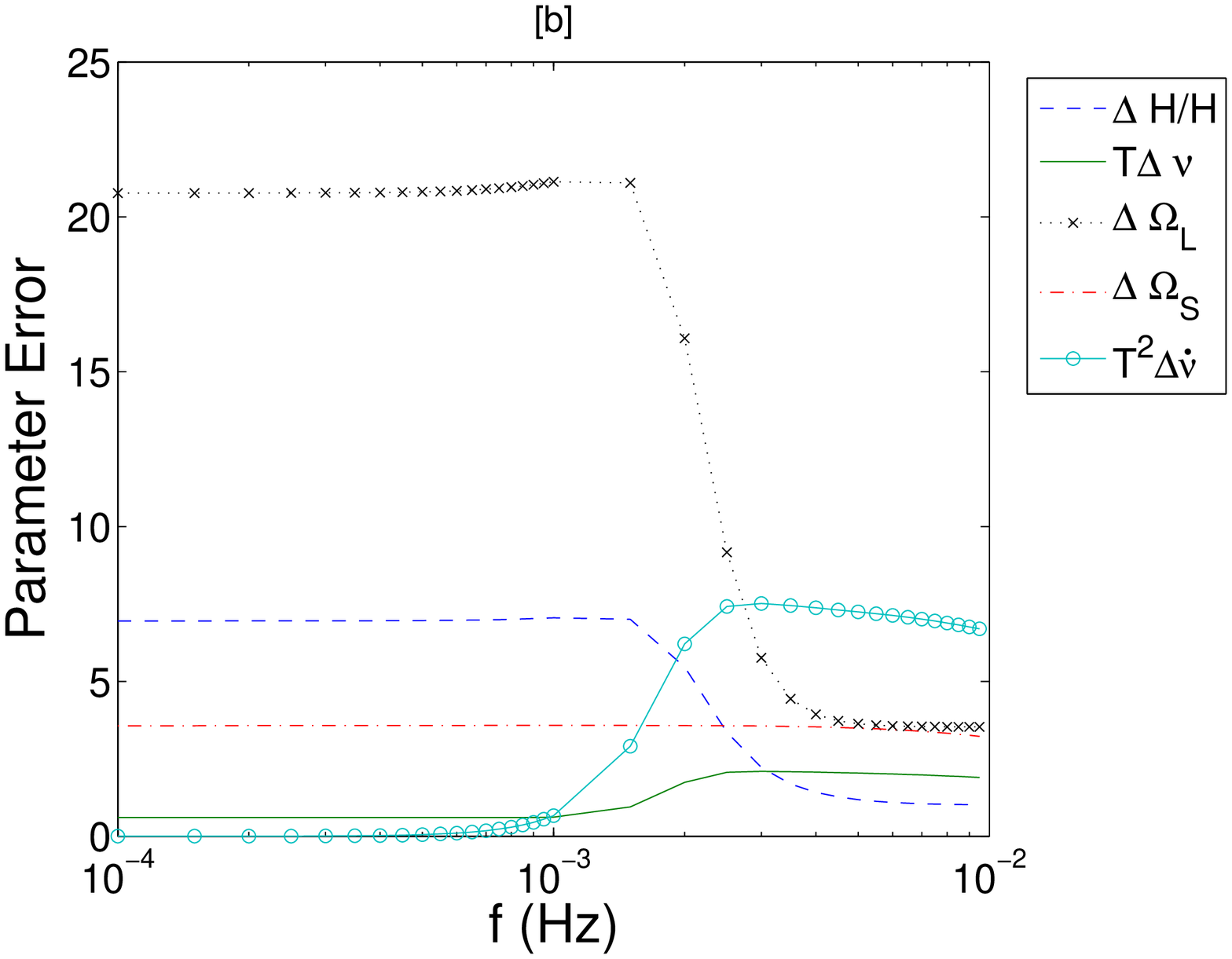,height=3.in,width=3.in}}
\centerline{\psfig{file=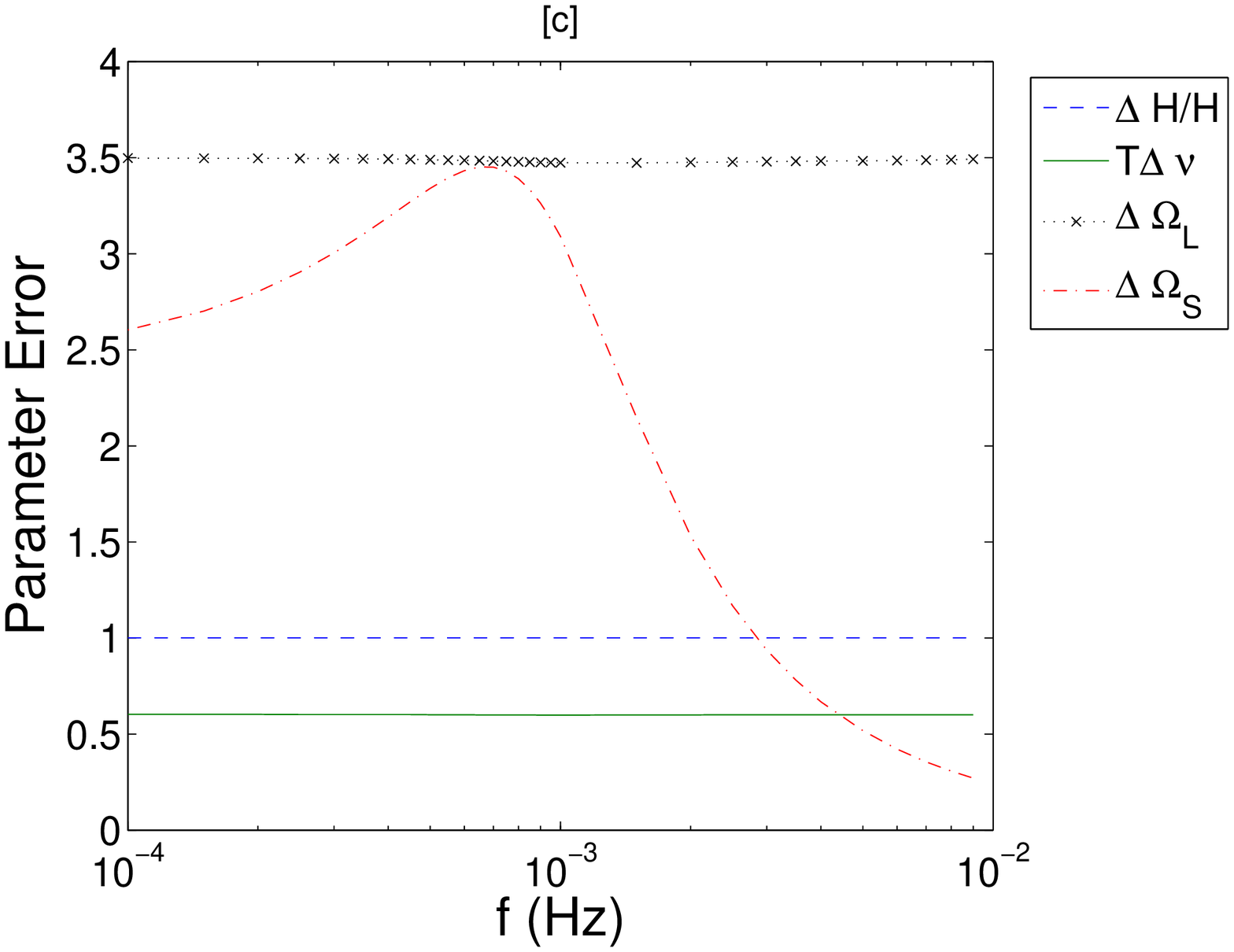,height=3.in,width=3.in}\psfig{file=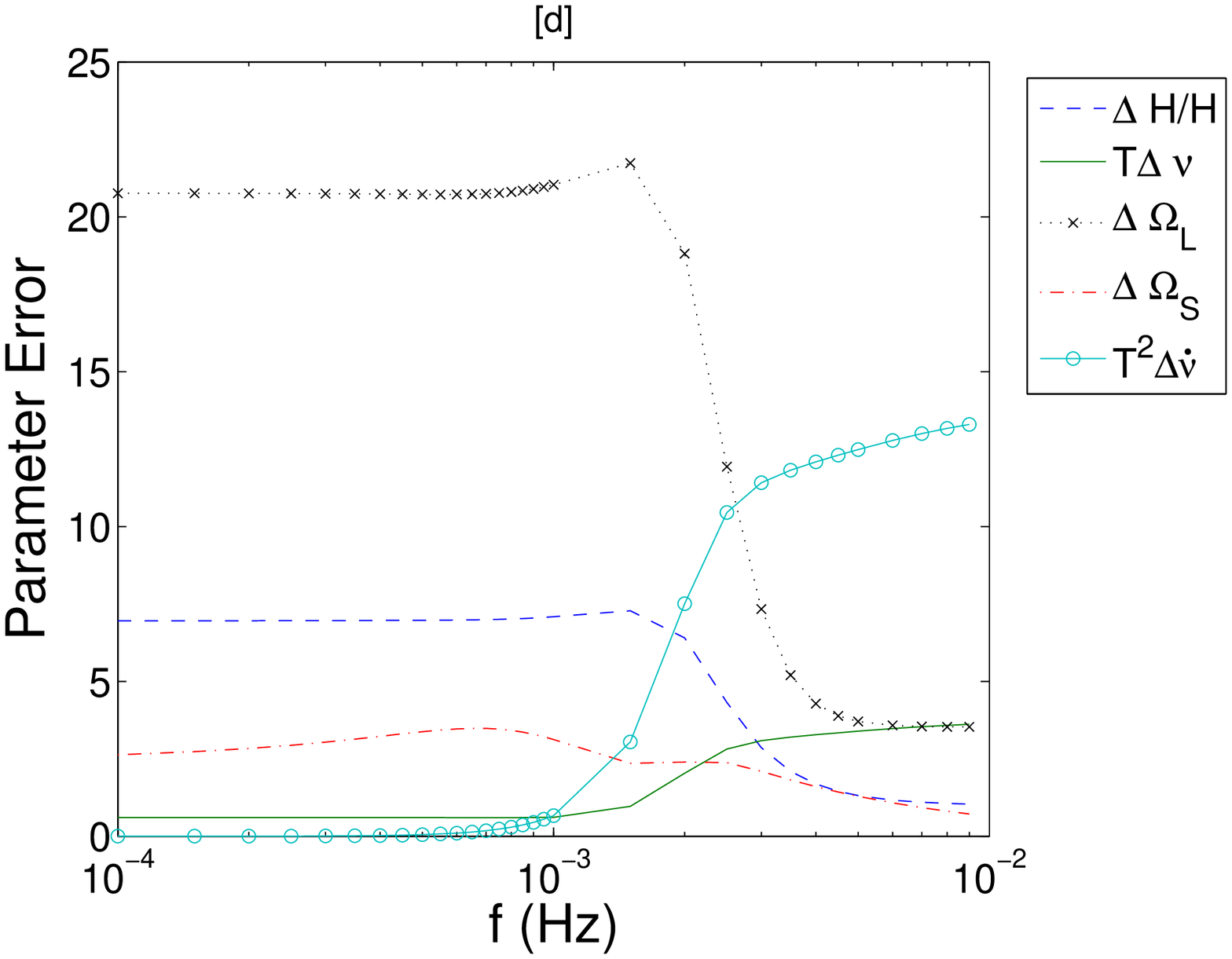,height=3.in,width=3.in}}
\caption{These figures show how the errors in the different signal
parameters behave as a function of the source frequency for the
following four cases (a) with no Doppler-phase modulation and no chirp, 
(b) with chirp only, but no
Doppler-phase modulation, (c) with Doppler-phase modulation, but no chirp,
and (d) with both Doppler-phase modulation and chirp included.
The sky-position assumed is: $\theta=\pi/2$, $\phi=\pi/2$; also,
$\iota=\pi/4,~\psi=\pi/3$.
} \label{freqplots}
\end{figure}

Figure \ref{freqplots3} shows that for binaries with
${\cal M}_c \simeq 5.3~M_\odot$, the error in the estimation of its chirp is
at least as large as the value of the chirp itself (for an SNR of 10)
for $\nu \simlt 2.5$ mHz. As shown in Fig. \ref{freqplots}, this is very close
to the frequency where the inclusion of the chirp starts hurting the estimation
of the frequency. On its own merit, this would argue for the dropping of
the $\dot{\nu}$ parameter from the search templates (i.e., setting
$\vt^0 = 0$)
for ${\cal M}_c \simgt 5.3~M_\odot$ and
$\nu \simlt 2.5$ mHz. However, the fitting factor plot
shows that doing so can vastly reduce the probability for detecting such signals
with $\nu \simgt 1$ mHz. We have thus, identified a region
in the parameter space where entertaining the prospects of a detection requires
taking a beating in estimating the source parameters. This adds an additional
layer of challenge in the quest for cleaning what is deemed as the resolvable
part of the confusion noise spectrum.

\section{\label{sec:freqDep}Frequency Dependence of parameter errors}

\begin{figure}[!hbt]
\centerline{\psfig{file=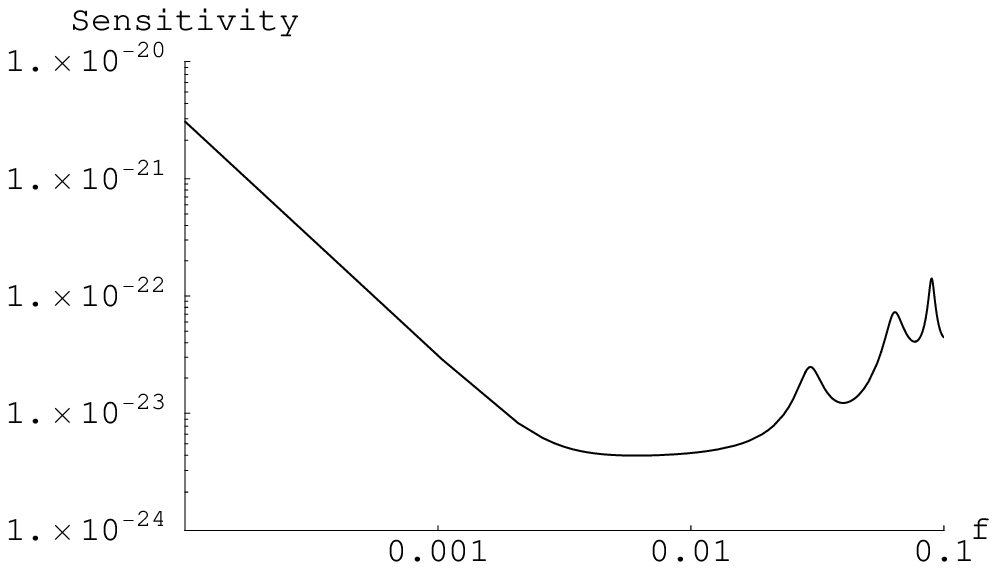,height=3.in,width=4.in}}
\centerline{\psfig{file=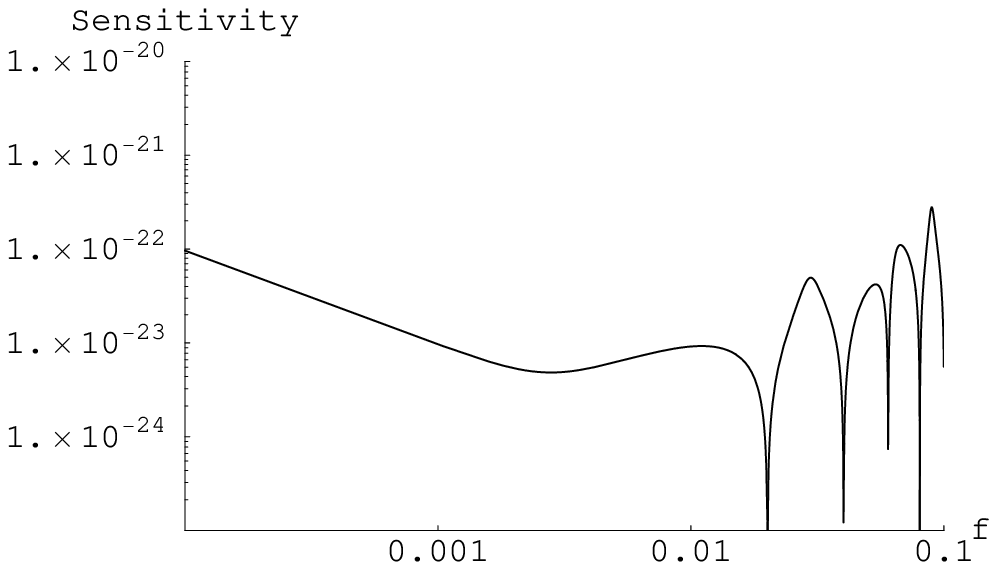,height=3.in,width=4.in} }
\caption{These are plots for the sensitivity of LISA as a function of 
the frequency (in Hz) for the first
and second generation TDIs, respectively.} \label{sensitivity}
\end{figure}

Imagine a one-parameter family of sources, which have everything else
identical, except for the emission frequency, $\nu$. The
parameter-estimation errors for such a family are not invariant
across LISA's bandwidth. This is because both the
Doppler-frequency shifting and the frequency evolution increase with $\nu$.
Additionally, the sensitivity of LISA itself varies across the band,
as shown in Fig. \ref{sensitivity}. For nearly monochromatic signals,
the latter effect gets essentially factored away when considering
{\it SNR-normalized} estimates.
Figure \ref{freqplots} shows the frequency dependence
of the parameter errors, normalized for an SNR of unity,
for four different cases:
Case (a) depicts these errors when a signal has no
Doppler-phase modulation, and no frequency evolution
(where the latter is unphysical).
Case (b) presents the same when there is frequency evolution
natural to the binary,
but no Doppler-phase modulation (such as if LISA were at the SSB).
Case (c) includes the effect of Doppler phase modulation but has the frequency
evolution (artificially) turned off. Finally, plot (d) has effects from both
the Doppler phase modulation and the frequency evolution.

It is interesting to note that in Fig. \ref{freqplots}a the error
in the estimate of each of the parameters is essentially independent
of the emission frequency.
The primary contribution to the frequency dependence of the Fisher
information matrix in Eq. (\ref{Fisher}) arises from LISA's own
sensitivity reflected in its noise PSD. The elements of this
matrix depend on $P^A(f)$, which is the same for both $A=1,2$.
However, when the parameter errors are obtained by
scaling this matrix by the template norm, as described in
Eq. (\ref{FisherNorm1}), the factor of $P^A(f)$ for near-monochromatic
signals gets canceled away, thus rendering the plots in
Fig. \ref{freqplots} mostly flat.

We next consider the effect of Doppler-phase modulation on the parameter
estimation accuracy. Comparing Figs. \ref{freqplots}a and
\ref{freqplots}c shows that the errors in all parameters are mostly
unaffected except for the error in sky-position.
That error actually worsens with frequency in the 0.1-1 mHz band, before
rapidly improving at higher frequencies. This is because at wavelengths longer
than 1 AU, the Doppler-phase modulation acts like noise on the sky-position
dependent $\sigma^A_i$ term in the phase of the signal in Eq. (\ref{signal}),
thus, affecting sky-position estimation. On the other hand, at high
frequencies, it does a better job at resolving a binary's direction
than do the beam-pattern-function induced modulations of the amplitude and
the phase.

The effect of the chirp can be understood by comparing Figs. \ref{freqplots}c
and \ref{freqplots}d (or, alternatively, Figs. \ref{freqplots}a and
\ref{freqplots}b): It worsens the estimation of the amplitude
and source orientation at low frequencies, i.e., for
$\nu \simlt 1.5$ mHz. It also reduces the accuracy of its own
estimate and that of the source
frequency at high frequencies, i.e., for $\nu \simgt 1.5$ mHz.
Here, one may wonder why
at frequencies below 1.5 mHz, the chirp's inclusion ruins the determination
of the source orientation, but not the frequency. The answer is that during
the course of a year one integrates through a very large number of cycles 
of a signal, which averages out the noise
in the estimation of the frequency arising from the chirp. The same would be
true about the orientation if one were to integrate for a large number of
LISA orbits (or years). Interestingly, at frequencies higher than 2 mHz,
while the chirp leaves the estimation of the source orientation unaffected,
it hurts the estimation of the source
frequency. This is because starting at around a milli-Hertz, the chirp 
acts as a source of noise in the determination of the frequency.

In summary, the general behavior of the SNR normalized parameter-estimation
errors as functions of $\nu$ can be understood as follows.
Primarily, there are two competing parts of the signal that provide
information about the source: The amplitude modulation and
the phase modulation. The amplitude is modulated differently
throughout the course of a year depending on the source location
and orientation. The amplitude modulations provide information about
these two source parameters in a manner that the SNR-normalized
error in them remains invariant across LISA's band 
(as can be inferred from Fig. \ref{freqplots} a). 
Whereas tracking the phase provides
information about the source location, source frequency, frequency
evolution and, hence, the chirp mass. The main difference between the two
modulations is that the phase modulation is highly dependent on the
emission frequency of the source, whereas amplitude modulation is not.
Since the amplitude modulation is less sensitive to the emission frequency,
its contribution to any parameter estimate is relatively unchanged over LISA's
band. Contrastingly, phase modulation provides less information at low
emission frequencies.
This is due to the relatively fewer cycles one obtains in a given observation
time for
low frequencies compared to higher ones. For the frequency range that LISA is
projected to observe, this translates into roughly a factor of
$10^3$ less cycles that would be obtained for the smallest
observable frequencies with respect to the largest ones.
However, as the frequency increases the contribution from the phase modulation
eventually surpasses that due to the amplitude modulations.

\subsection{\label{subsec:altInterp} Excluding frequency evolution from
the templates}

The preceding discussion shows that at any given LISA frequency
{\em searching for} the frequency evolution of a binary affects the
determination of one or the other parameter, except the sky position.
For a source with an initial frequency, $\nu \simgt 1.5$ mHz,
its inspiral hurts the error in the estimation of
$\nu$ by a factor of over 3.5 relative to that of sources
with $\nu \simlt 1$ mHz. We have verified that increasing the
integration time to 2 years reduces this relative error factor to about 2,
confirming what is shown in Ref. \cite{Takahashi:02}. 
This reduction may be insufficient for the resolution of the galactic binary 
confusion noise. This makes a case for a mission lifetime that is longer 
than 2 years. Otherwise, the only other option available for maintaining 
the parameter accuracies is to exclude the chirp parameter from the search
templates, since doing so improves the
SNR-normalized parameter estimates. This, however, has different consequences for
the parameter accuracies of binaries with different chirp masses.

The fitting-factor plots in Fig. \ref{omegadotmismatch} show that
for the lowest astrophysically interesting chirp mass,
${\cal M}_c\simeq 0.3M_\odot$, such an exclusion has little effect
on the SNR until about 4 mHz,
which happens to be just below the high-frequency-end of the confusion noise
arising from extreme mass-ratio inspiral (EMRI) sources. The effect on
the estimates of the other parameters is a systematic bias by an
amount that depends linearly on $T^2\dot{\nu}$ \cite{BoseRoganPrep1},
which is about 0.1 (cycles) for $T=1$year and at $\nu = 4$ mHz. While this introduces
a negligible bias in the estimates of all parameters, it has the advantage
of reducing the level of the random estimation errors to the same level
as that shown in Fig. \ref{freqplots}a.
At frequencies as high as even 6 mHz, the SNR drop due to such an
exclusion would be more than 30\%, thus, jeopardizing both the prospects
of detection and the adequate parameter resolution required for cleaning the
WD-WD confusion noise.

On the other hand, note that for a
binary with ${\cal M}_c \geq 5.3M_\odot$, such an exclusion causes the same
fractional drop in the SNR at frequencies as low as 1.5 mHz, which is where
the parameter estimates start getting affected (see Fig. \ref{freqplots}b
or \ref{freqplots}d). Thus, for binaries with ${\cal M}_c \geq 5.3M_\odot$,
turning off the chirp parameter in the search templates is unlikely to improve
the parameter estimation for $\nu \geq 1.5$ mHz.
Among the sources most affected from this ``chirp search'' problem are 
galactic low-mass binaries containing a black hole.

\section{\label{sec:errorSkyMaps}Sky-position dependence of parameter errors}

In this section, we study the behavior of the parameter errors as functions
of the direction to a binary. We illustrate the sky-position dependence of 
the SNR-normalized parameter errors for the same source polarization state 
and frequency as the one chosen in Sec. \ref{sec:freqDep} for the 
frequency plots in Fig. \ref{freqplots}. The source frequency choice, viz., 
$\nu = 3$ mHz is interesting for multiple reasons: As observed above, it 
is close to
the high-frequency-end of the unresolvable WD-WD confusion noise. Figure
\ref{freqplots}d shows that it is also in the region of transition in the
behavior of the parameter errors as functions of frequency. All the same,
it is not too small for the Doppler-phase modulation to have a negligible
effect on the parameter accuracies. 

When studying these errors
as functions of frequency in Fig. \ref{freqplots}, we had fixed the
sky position of the source. For this reason, while
computing the SNR-normalized errors, we did not have to worry about the
possibility that the SNR may vary with the sky position. We begin
by addressing that possibility below. The dependence of the SNR on the
polarization state can be studied similarly.

\begin{figure}[!hbt]
\centerline{\psfig{file=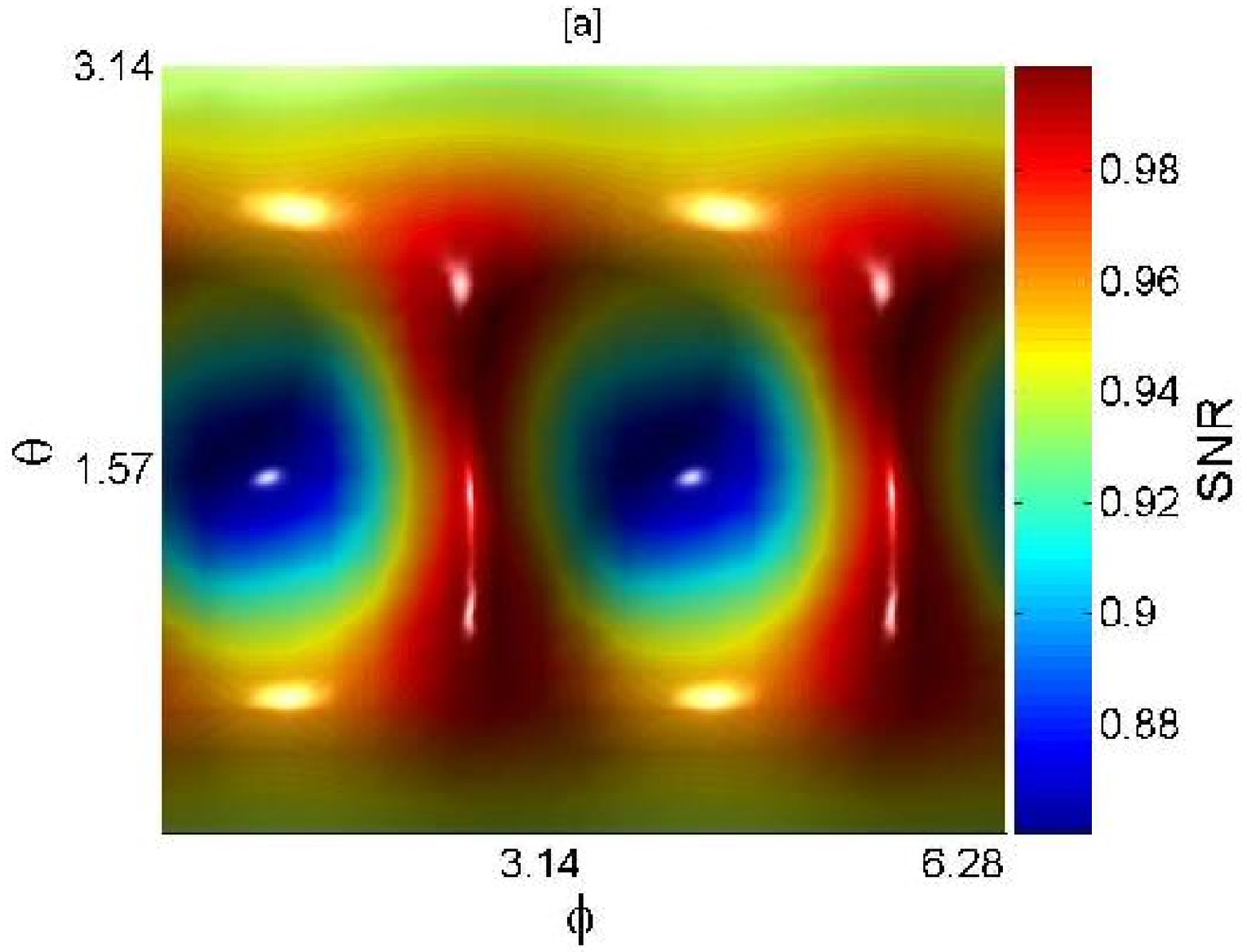,height=3.5in,width=3.5in}\psfig{file=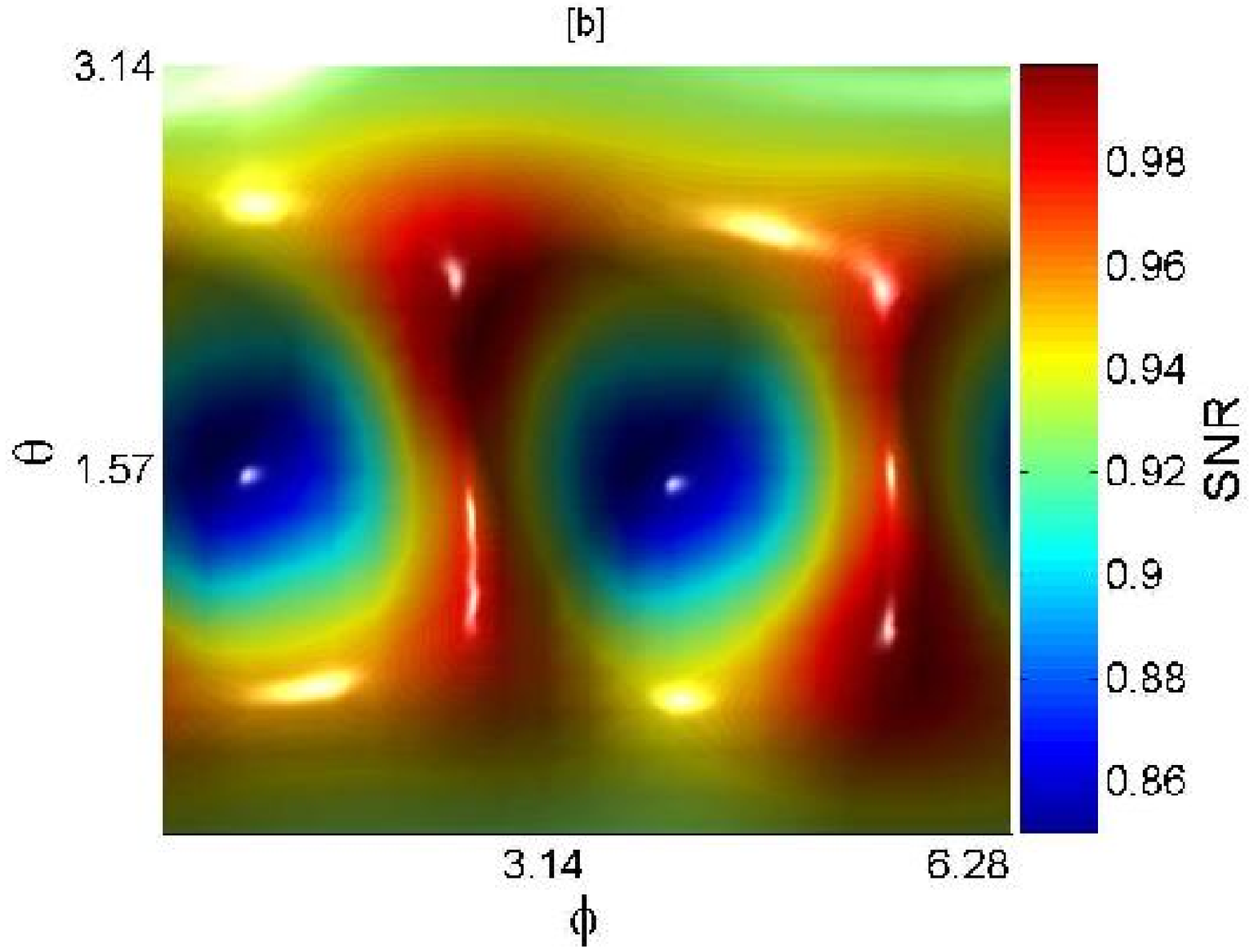,height=3.5in,width=3.5in}}
\centerline{\psfig{file=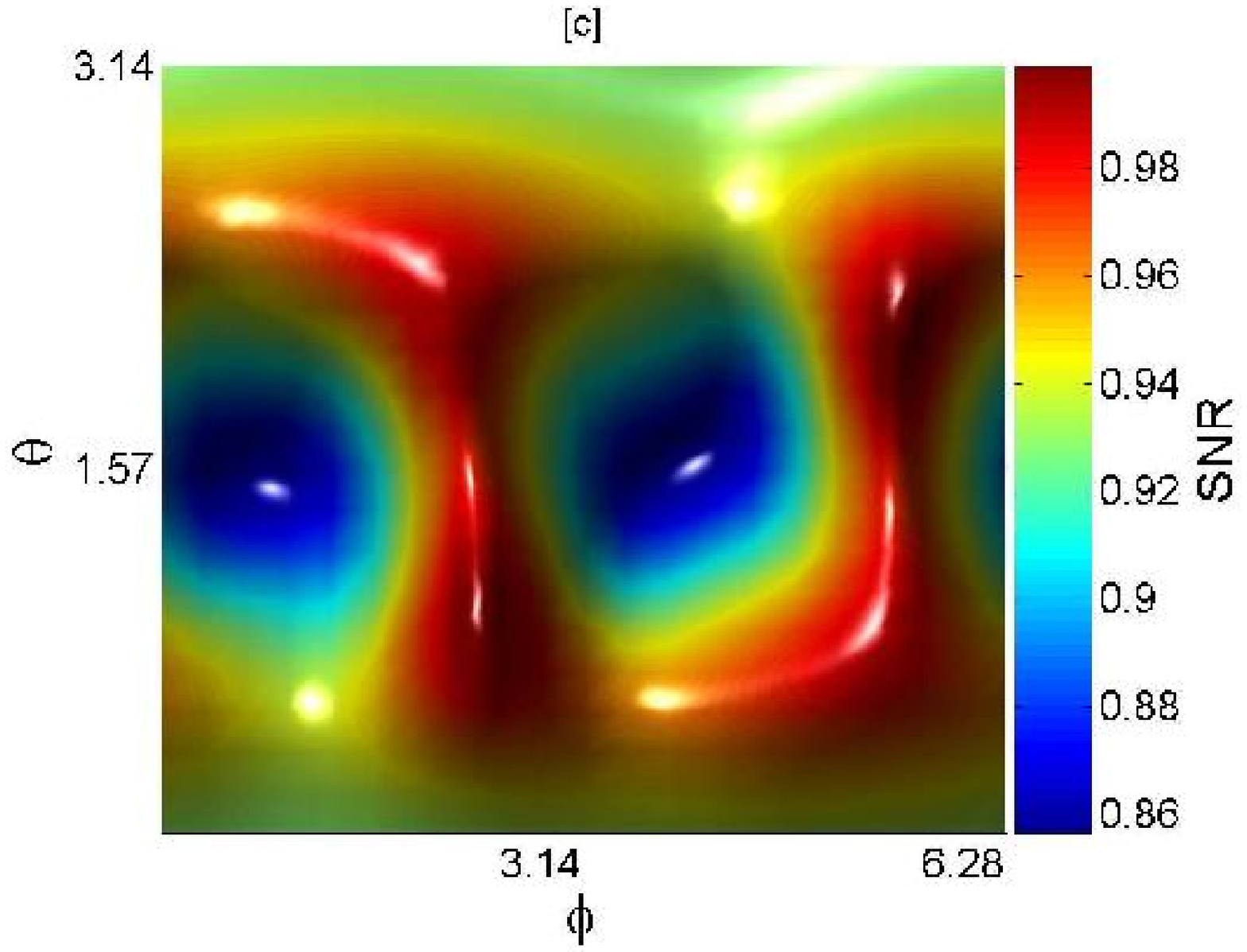,height=3.5in,width=3.5in}\psfig{file=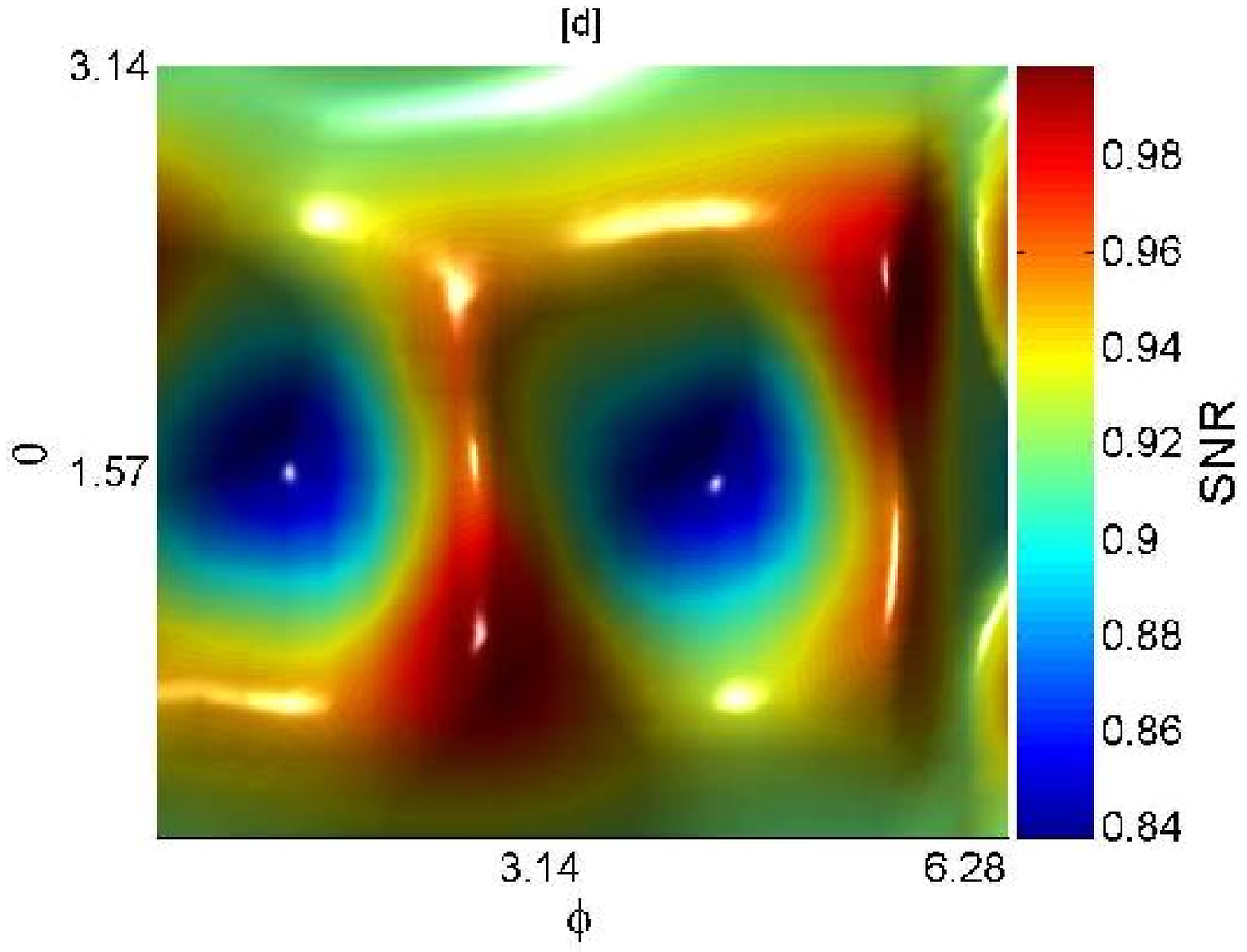,height=3.5in,width=3.5in}}
\caption{These are plots of the template-norm-squared for the
same four cases of LISA motion and source frequency evolution considered in
Fig. \ref{freqplots}. The plot for each case is normalized to have a maximum
value of unity. Both $\theta$ and $\phi$ are given in radians, with the origin,
$(0,0)$, in the bottom left corner of each plot. 
Note that this is the first of a series of {\em colored} sky-plots; 
gresyscale prints of these 
plots can give a misleading impression of the color values.}
\label{SNRvsInfo}
\end{figure}

\begin{figure}[!hbt]
\centerline{\psfig{file=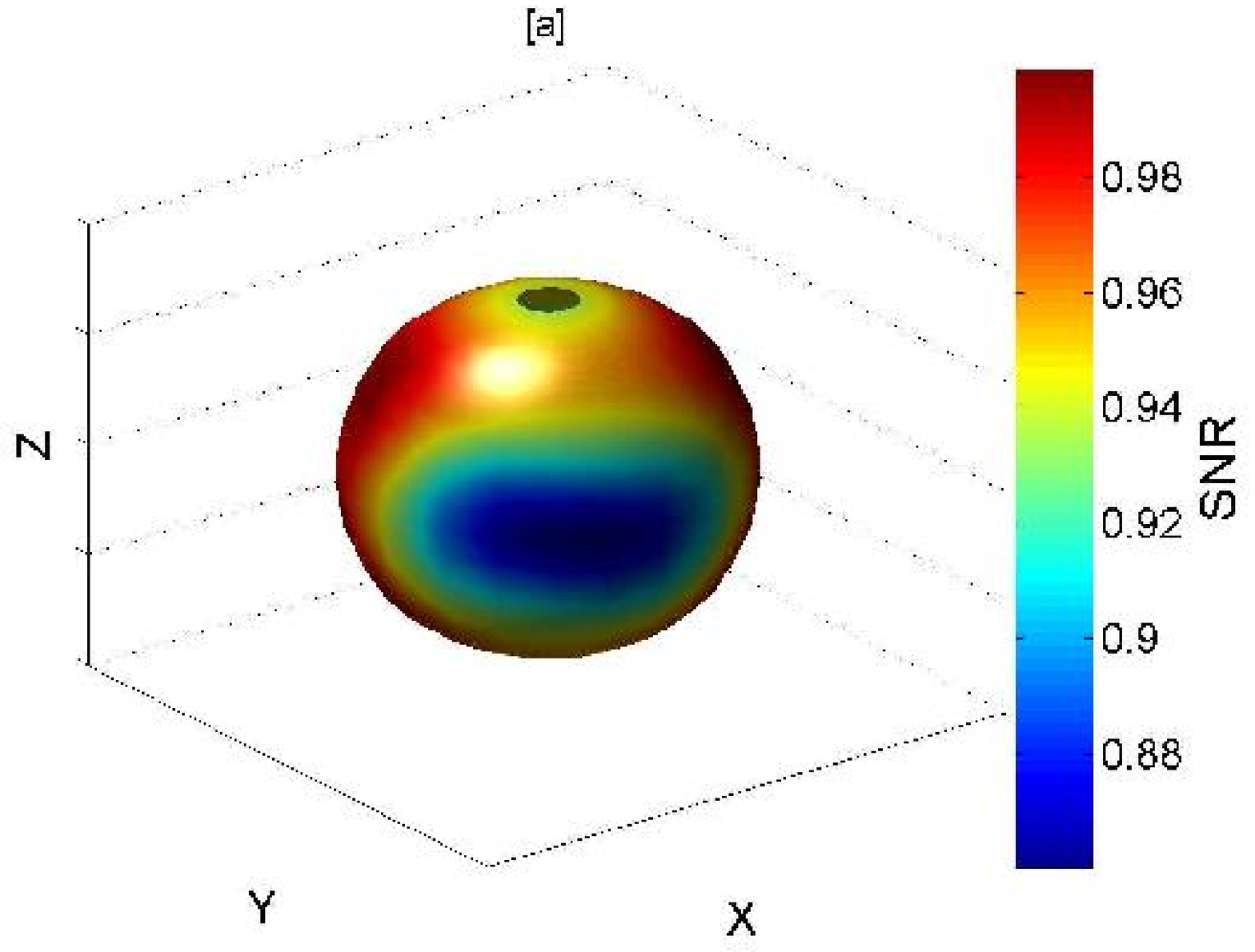,height=3.5in,width=3.5in}\psfig{file=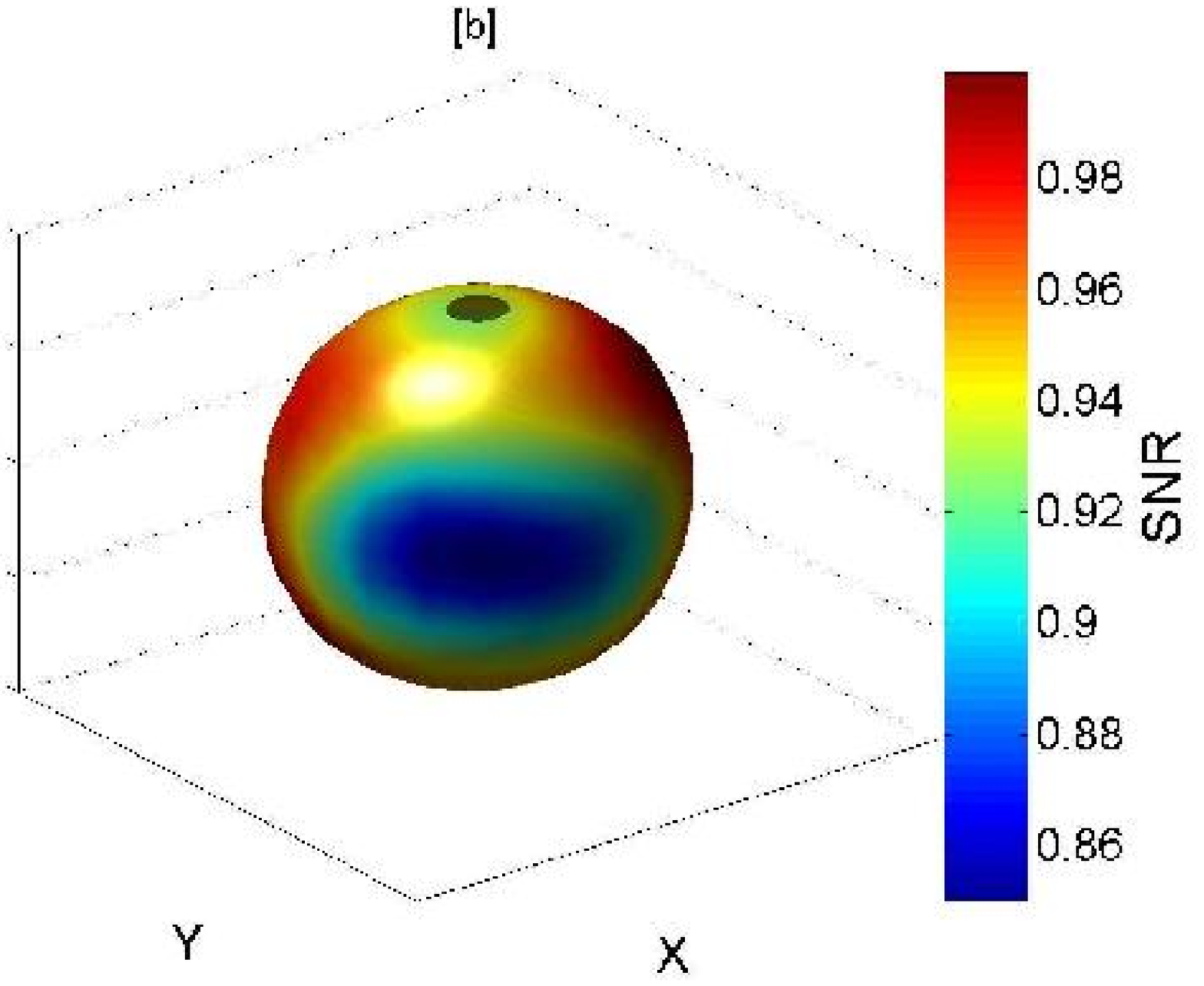,height=3.5in,width=3.5in}}
\centerline{\psfig{file=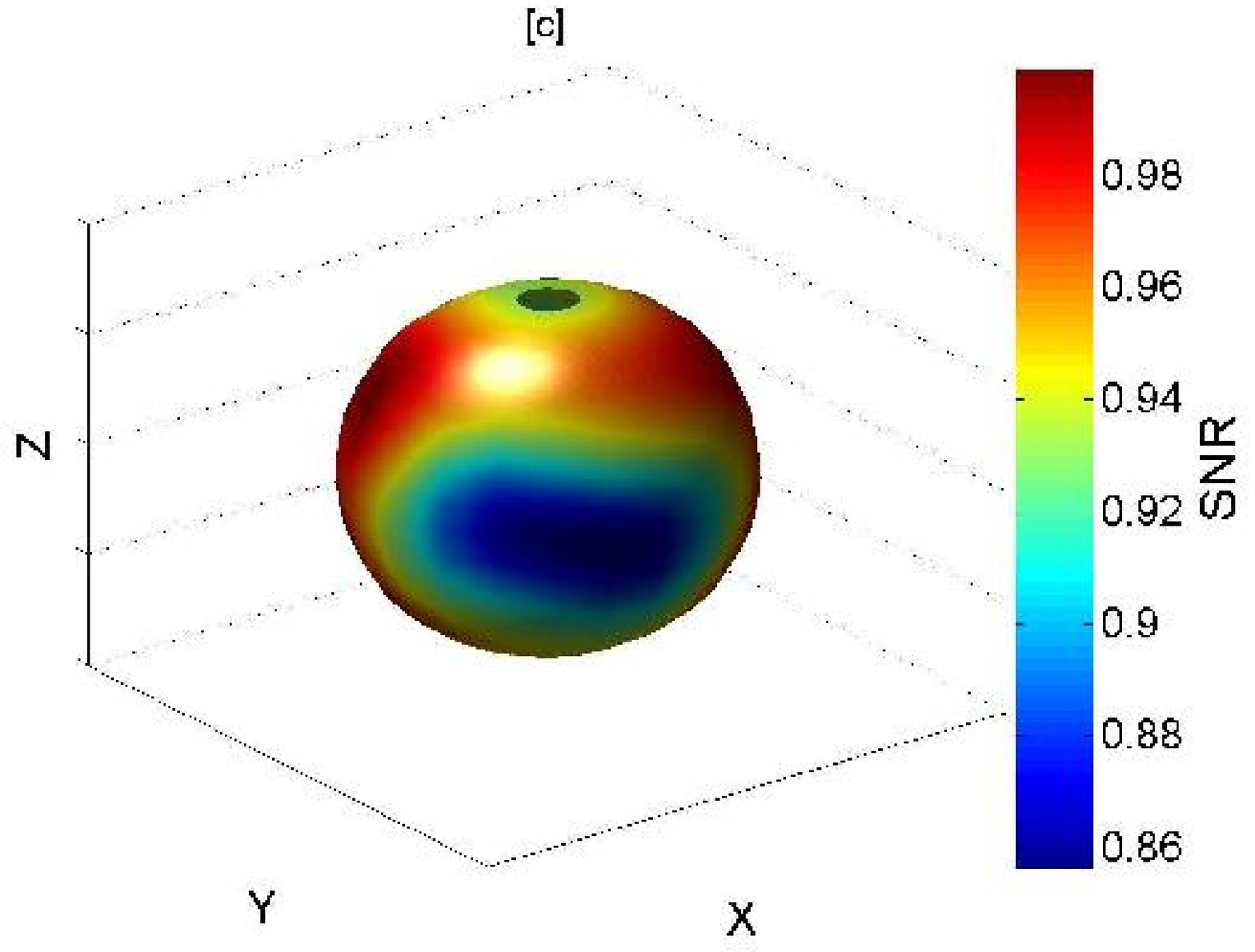,height=3.5in,width=3.5in}\psfig{file=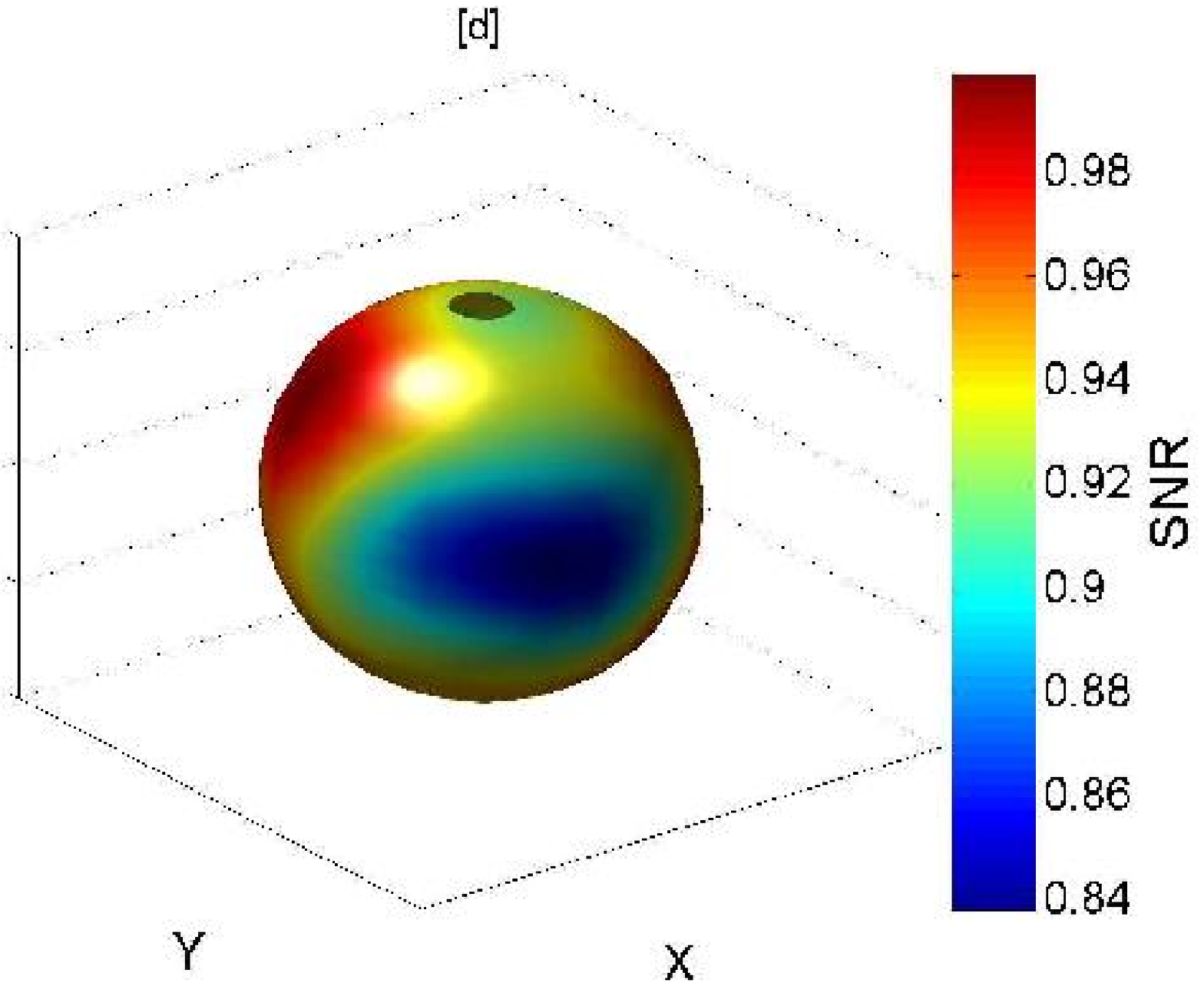,height=3.5in,width=3.5in}}
\caption{
Spherical renditions of the template-norm-squared plots shown 
in Fig. \ref{SNRvsInfo}.
} \label{SNRSphereVsInfo}
\end{figure}

\subsection{\label{subsec:snrSky}Sky-position dependence of the SNR}

The SNR normalization performed on the Fisher information matrix in
Eq. \ref{FisherNorm1}
is useful in that the parameter errors computed from such a matrix can be
easily scaled for a signal of any SNR using the scaling factors
given in Table \ref{tab:snrScales}. In this context, it is important to
note that the SNR of a standard candle is not uniform across the sky owing
to the non-uniform sensitivity of LISA to different sky positions. We plot
this quantity for four different cases in Figs. \ref{SNRvsInfo}
and \ref{SNRSphereVsInfo}, where the latter set is the 3D rendition of the
former set. These plots, called ``sky plots'',
are normalized so that the maximum value attained
in each case is unity. These four cases are identical
to those considered for studying the frequency dependence of
parameter errors in Figs. \ref{freqplots}. As in that study, here too
Fig. \ref{SNRvsInfo}a (or, equivalently,
Fig. \ref{SNRSphereVsInfo}a) is for an artificial signal, with
$\dot{\nu}$ set to zero), when compared with Fig. \ref{SNRvsInfo}b,
it allows one to graphically assess how
useful or detrimental the presence of frequency evolution is in
detecting a signal.
Similarly for Fig. \ref{SNRvsInfo}c, when compared with 
Fig. \ref{SNRvsInfo}d.
Note that as per the definitions of cases (a) and (b),
Figs. \ref{SNRvsInfo}a and \ref{SNRvsInfo}b have the
Doppler-phase modulation turned off.

\subsection{\label{subsec:skyDep}Explaining the sky-position dependence}

\begin{figure}[!hbt]
\centerline{\psfig{file=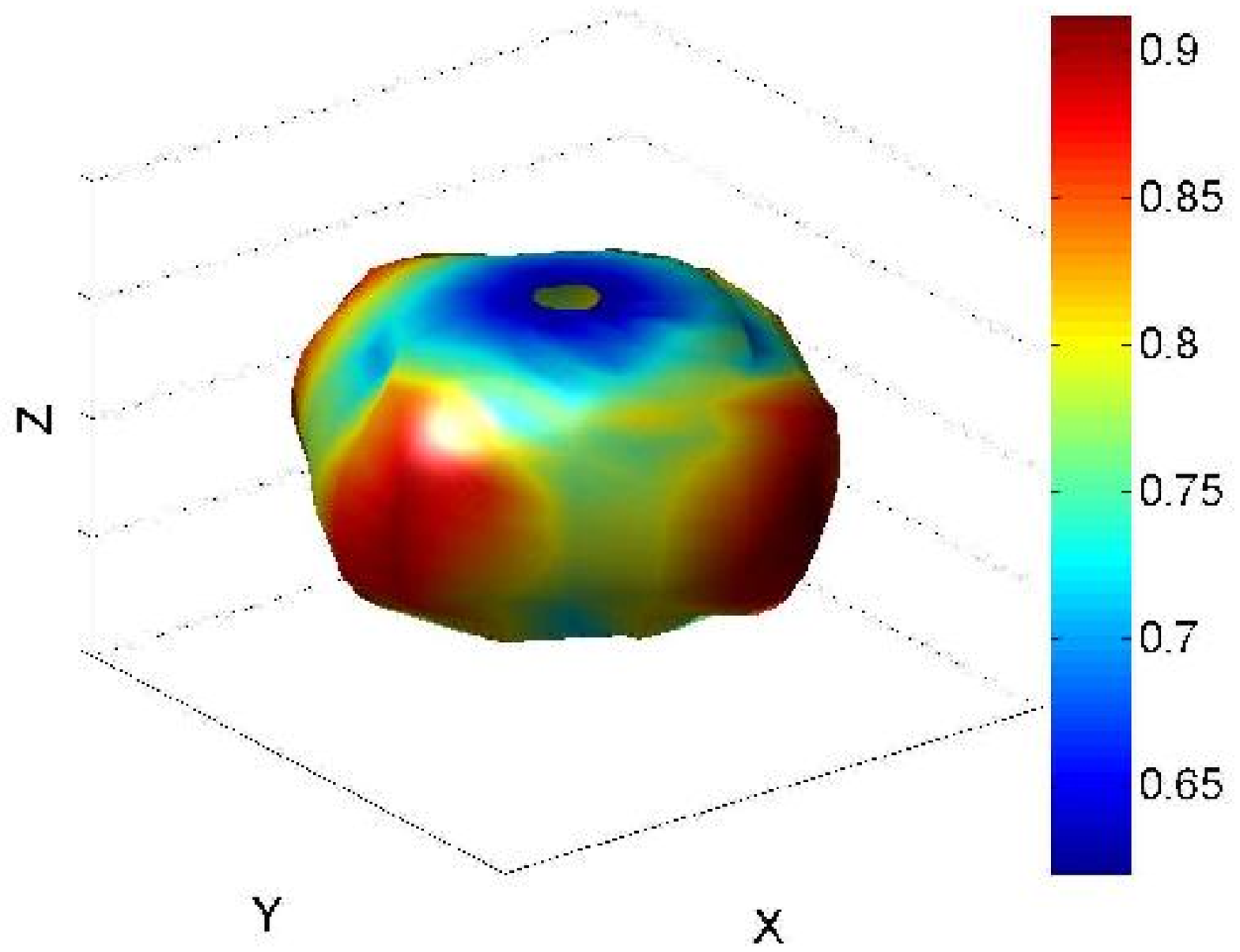,height=2.5in,width=2.5in}\psfig{file=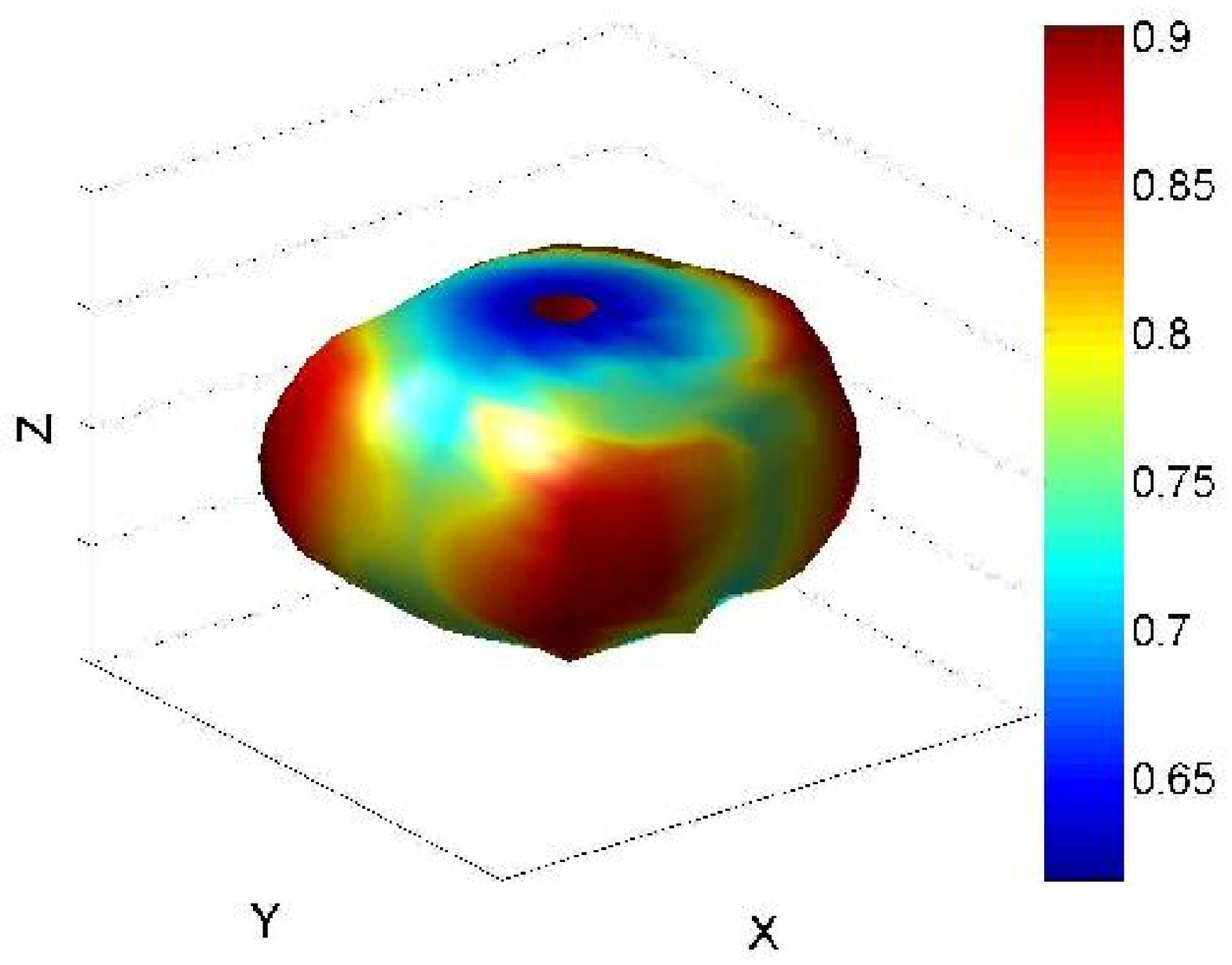,height=2.5in,width=2.5in}\psfig{file=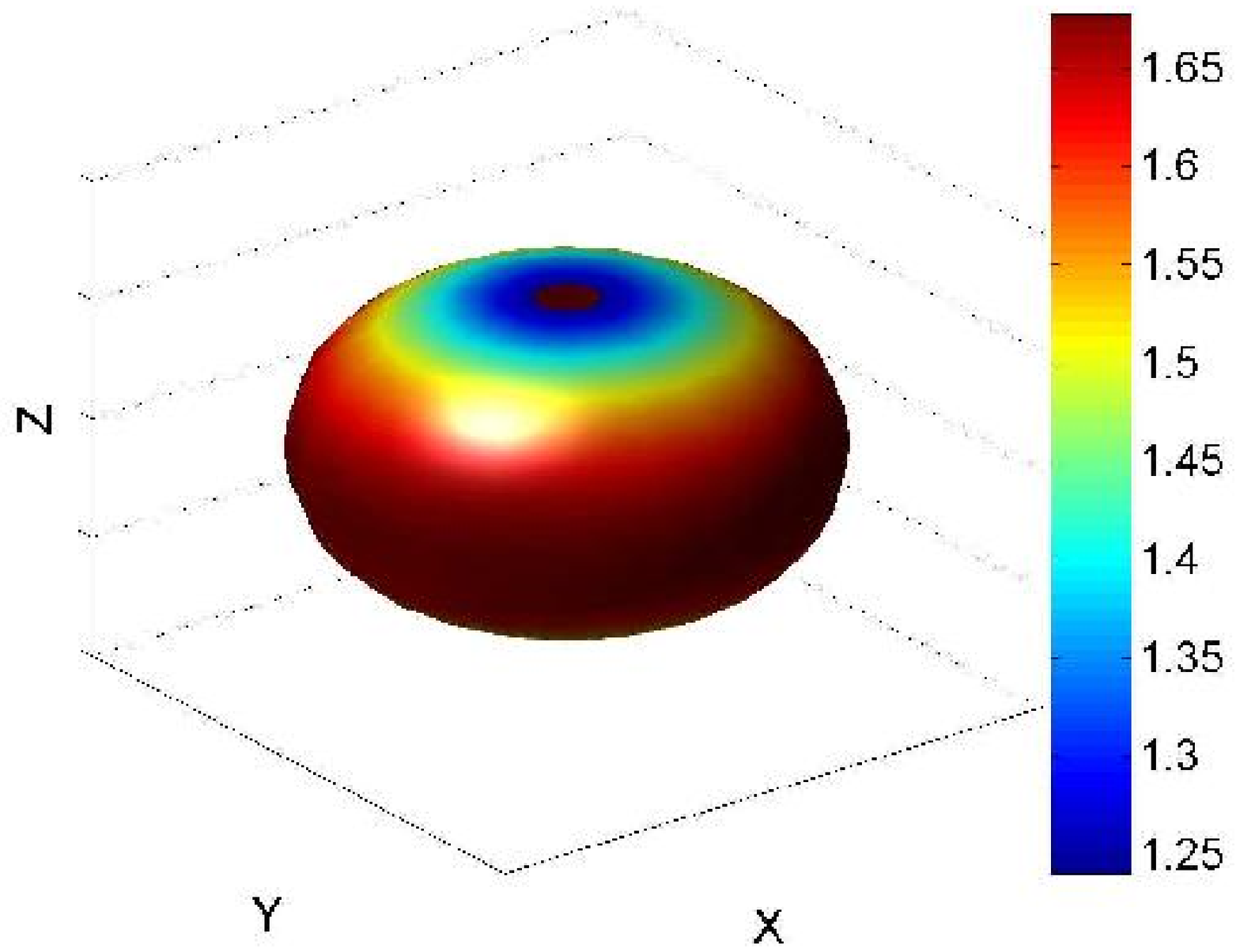,height=2.5in,width=2.5in}}
\caption{The template-norm-squared for the Michelson variables as functions of 
sky position:
for Michelson variable I as a function of sky position (left), II. for Michelson variable II as a function of sky position (center), III. for the network comprising both Michelson variables (right). All plots are presented for the following source parameters:
 $\{\iota=\pi/4, \psi=\pi/3, \nu=3~{\rm mHz}\}$}
\label{fig:michelsonSNR}
\end{figure}

\begin{figure}[!hbt]
\centerline{\psfig{file=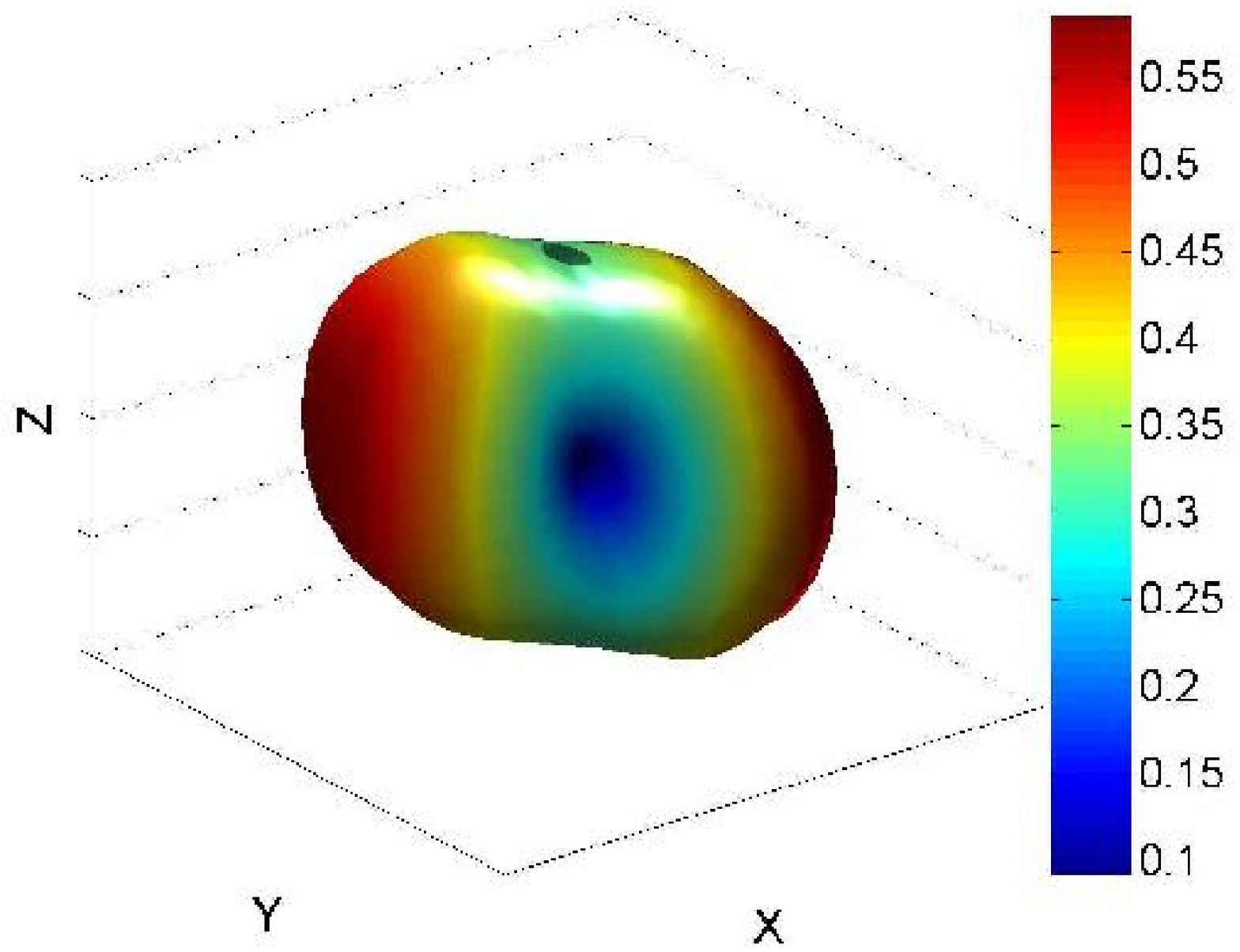,height=2.5in,width=2.5in}\psfig{file=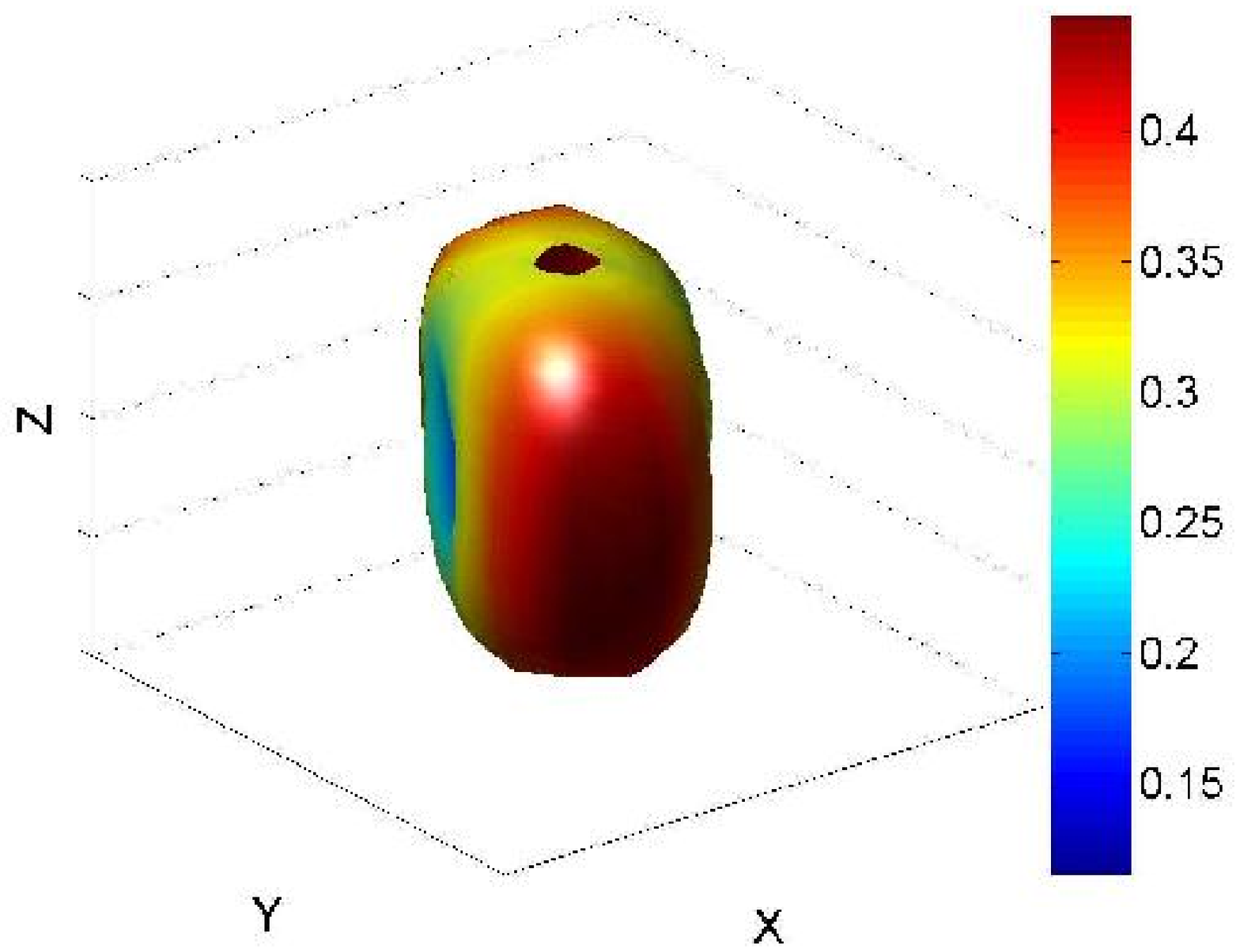,height=2.5in,width=2.5in}\psfig{file=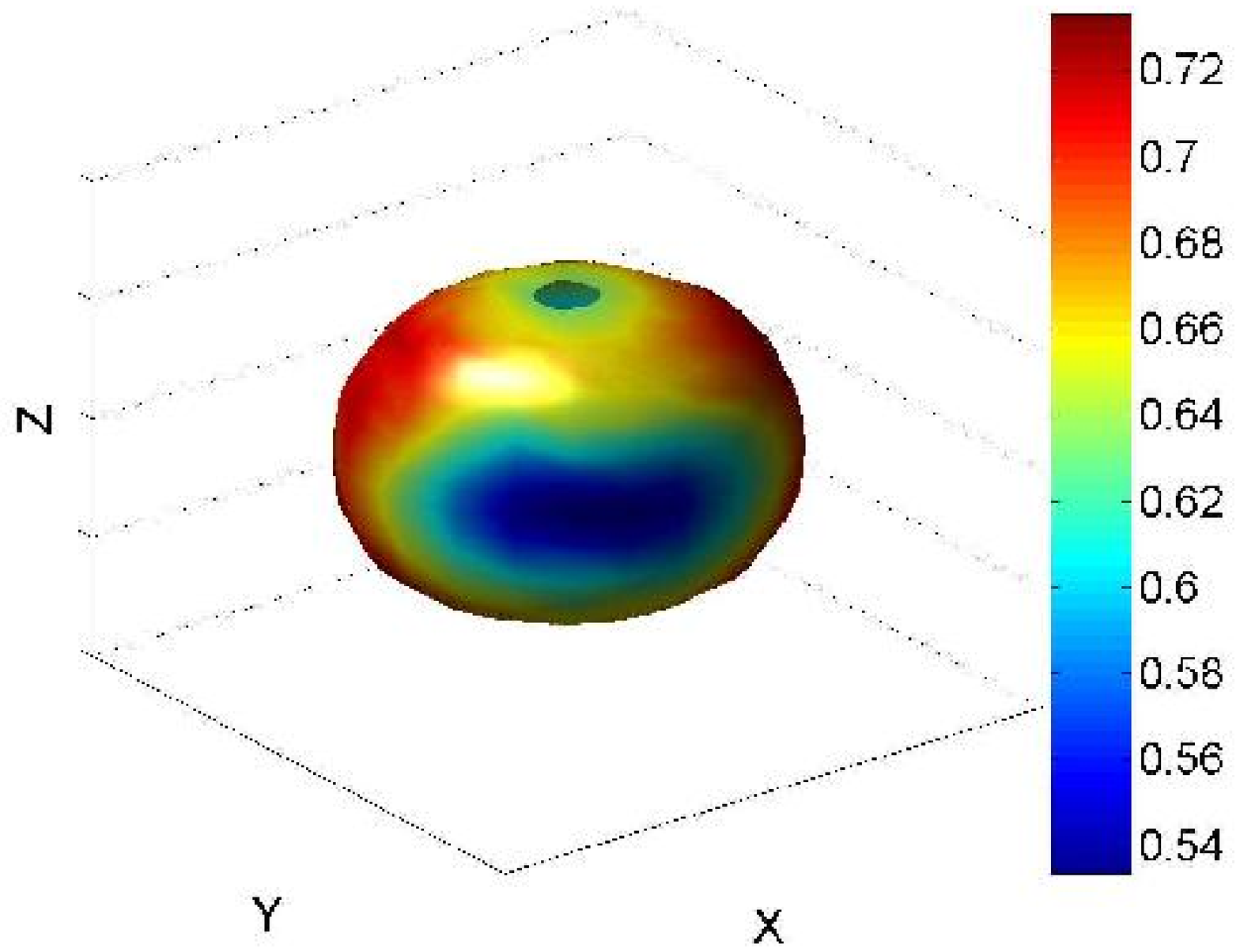,height=2.5in,width=2.5in}}
\caption{The template-norm-squared for the $\bar{A}$ and $\bar{E}$ TDI variables
as functions of sky position:
I. for the $\bar{A}$ TDI data combination (left), II. for
the $\bar{E}$ TDI data combination as a function of sky position (center),
and III. for the the network comprising both $\bar{A}$ and $\bar{E}$
TDI data combinations. All plots are presented for the following source parameters:
$\{\iota=\pi/4, \psi=\pi/3, \nu=3~{\rm mHz}\}$.}
\label{fig:ourSNR}
\end{figure}

Given that LISA's orbital motion is confined to a nearly circular orbit on
the ecliptic, one may expect the SNR plots to be axisymmetric. This
is indeed the case for the network comprising the two Michelson variables,
as can be seen in Fig. \ref{fig:michelsonSNR}, where we limit our focus
to their geometric sensitivity and momentarily ignore the fact that they
will be severely limited by the laser frequency-fluctuation noise.
Figures \ref{fig:michelsonSNR}a
and \ref{fig:michelsonSNR}b reveal that although the SNR for neither of the
Michelson variables is axisymmetric, their quadrature combination is so. 
For comparison, we present the analogous plots for the $\bar{A}$ and 
$\bar{E}$ TDI variables in Figs. \ref{fig:ourSNR}.
We normalize all these plots relative to the maximum value attained among
the template norms of the individual Michelson and $\bar{A}$-$\bar{E}$ variables.

Interestingly, however, the SNRs
for the $\bar{A}$ and $\bar{E}$ TDI variables, plotted in
Fig. \ref{fig:ourSNR},
are more dumbbell-shaped than axisymmetric, where the pattern depends on the
initial orientation of LISA. Also, since there is a
slight difference between the overall sensitivities of the two
variables, their quadrature combination is more of an inflated version
of the $\bar{A}$ variable. As one would expect, it also resembles well the
SNR of Fig. \ref{SNRSphereVsInfo}a.

\begin{figure}[!hbt]
\centerline{\psfig{file=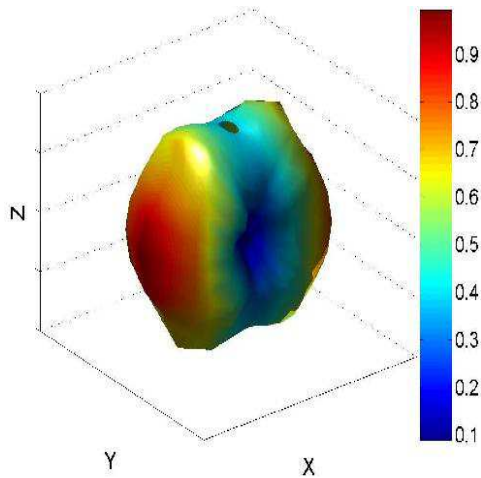,height=2.5in,width=2.5in}}
\caption{
A single arm's template-norm-squared. All plots are presented using the following parameters:
 $\{\iota=\pi/4, \psi=\pi/3, \nu=3~{\rm mHz}\}$.}
\label{fig:singleSNR}
\end{figure}

Figure \ref{fig:singleSNR} illustrates the sensitivity of a single arm to all sky positions. What is evident from this figure is its 
striking similarity to the $\bar{A}$ TDI data combination. The features of this plot are well understood geometrically. 
For a single arm, spinning and orbiting the SSB, there exists a periodicity of $\pi$ radians under axial rotations.
However, this figure shows that 
there are two distinct and orthogonal directions, and their antipodes, 
that stand out. These locations correspond to the 
global maxima and minima in the sensitivity of
this arm to impinging gravitational waves. These very same extrema are 
present in the SNR plots as well.
To understand the origin of these global extrema, the orbital motion of LISA's centroid is not as important as the spin of LISA's individual arms. Therefore,
consider how a single arm spins during the course of a year. Let it begin in an orientation that is parallel to the ecliptic plane. In this orientation, a low-frequency source that is in a direction perpendicular to the arm will 
create the maximum strain in that arm. Conversely, when such a source is
in a direction along the arm, the strain caused in that arm is zero. Let us 
call these directions 1 and 2, respectively. With the exception of their antipodes, there is no other orbital position that will attain such extrema 
in the sensitivity for that particular arm during its orbit. 
As the arm rotates out of the ecliptic plane, the sensitivity to direction 1 will begin to degrade. However, 
because direction 1 will never be parallel to the same arm, it will never attain the same minima
that direction 2 attained initially and, vice versa. Therefore, 
the orbital position where these global extrema are obtained is forever imprinted upon the integrated 
results regardless of the duration of observation.

\subsection{\label{sec:snrSky}Parameter errors}

\begin{figure}[!hbt]
\centerline{\psfig{file=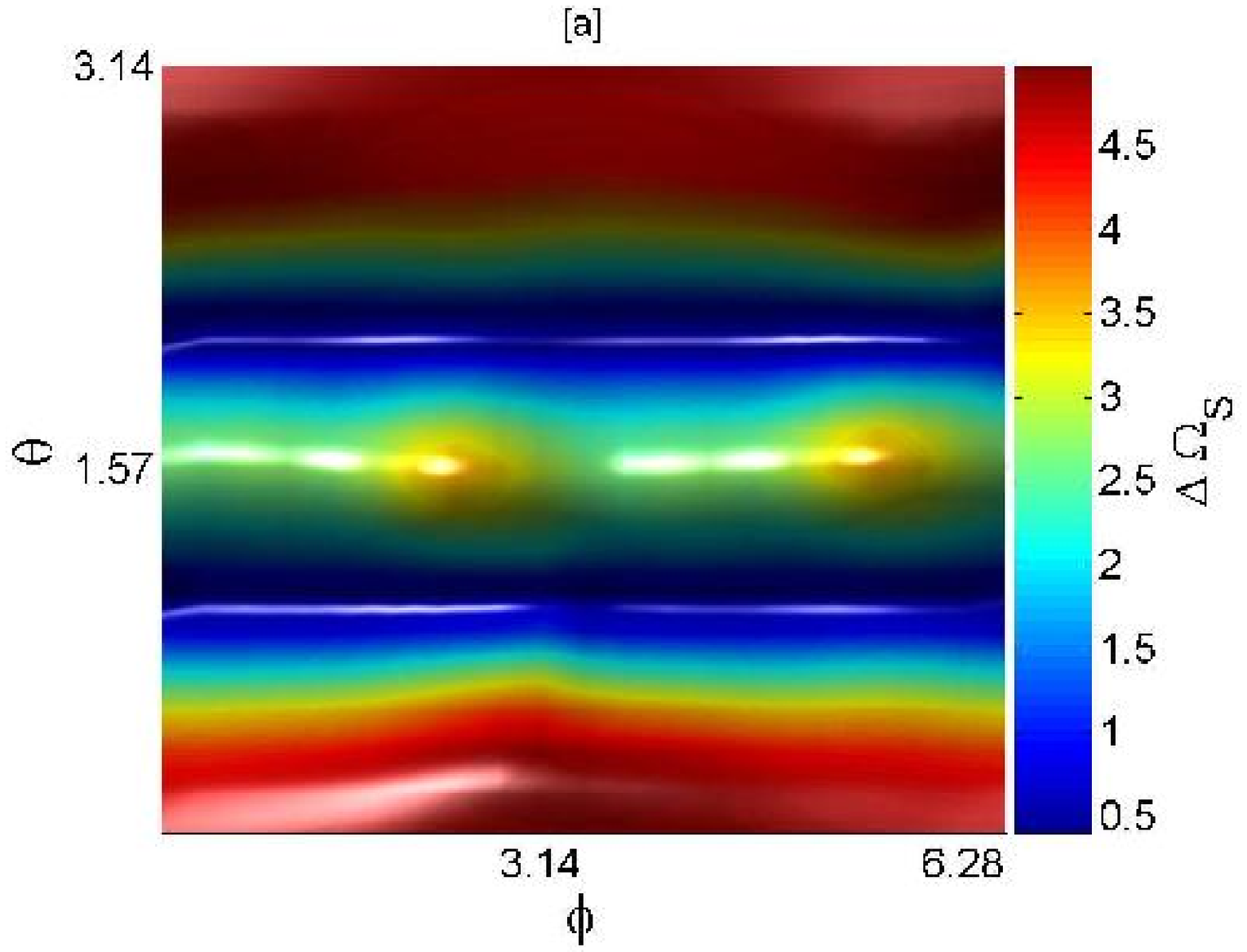,height=3.5in,width=3.5in}\psfig{file=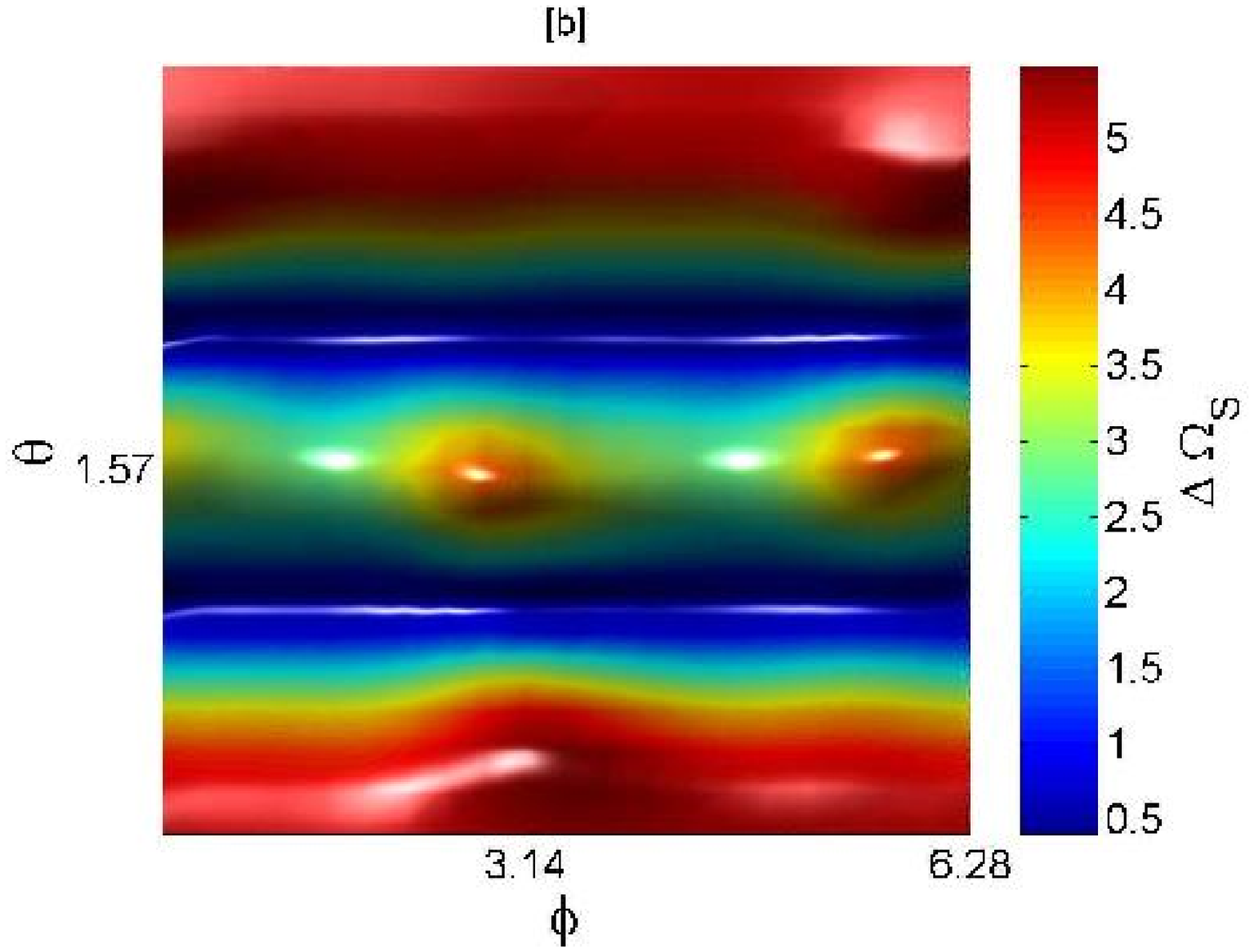,height=3.5in,width=3.5in}}
\centerline{\psfig{file=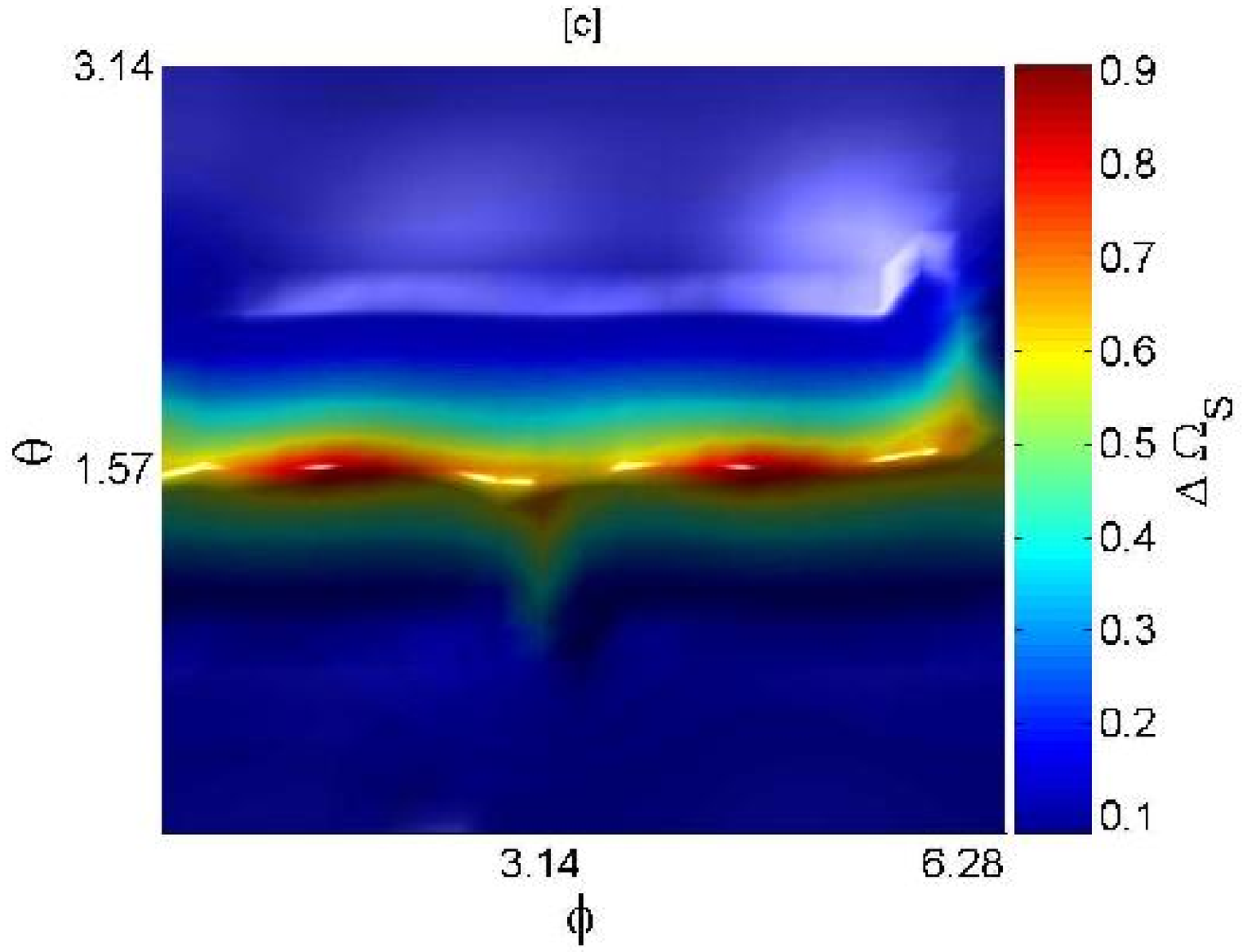,height=3.5in,width=3.5in}\psfig{file=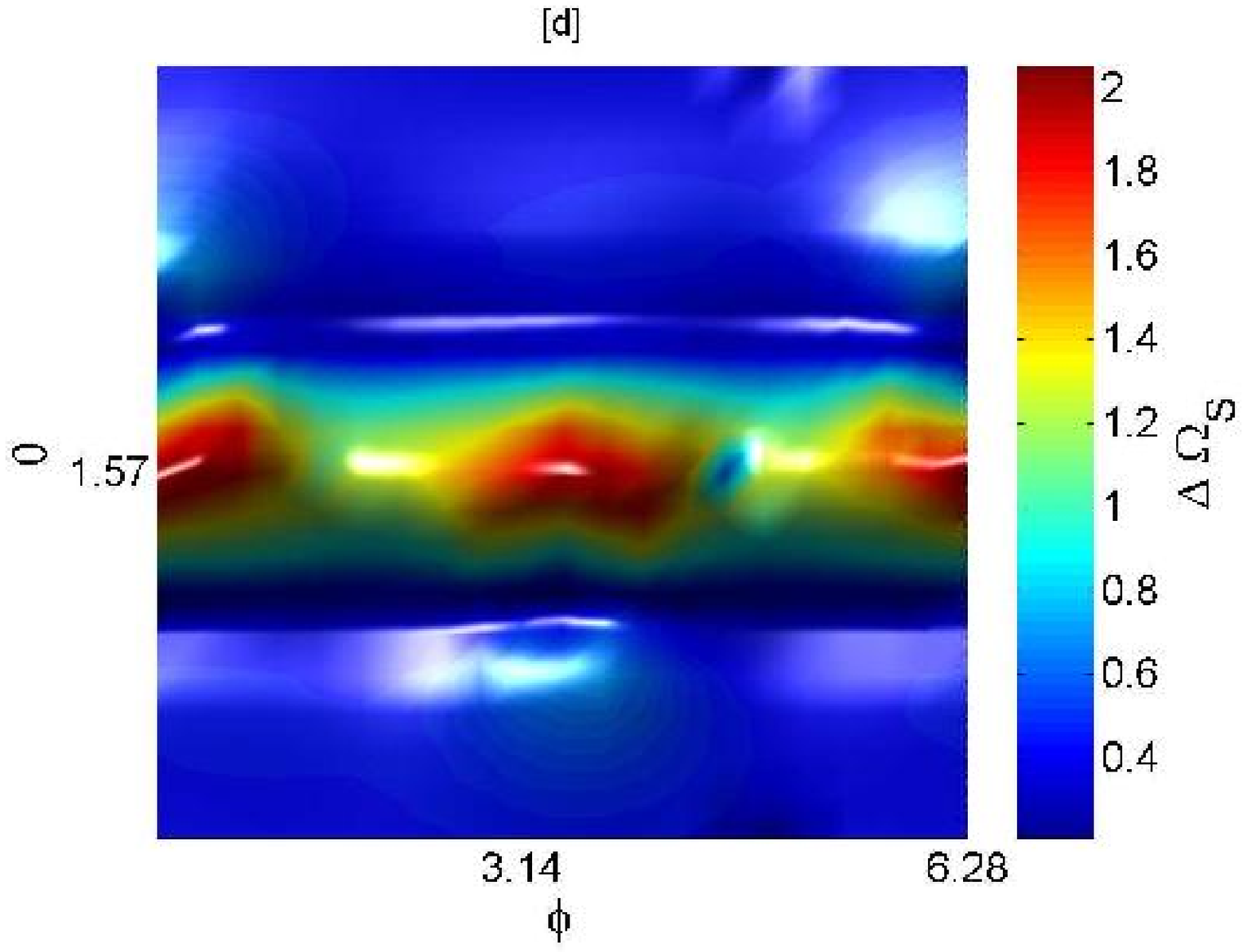,height=3.5in,width=3.5in} }
\caption{These are plots of the errors associated with determining
the sky position measured in terms of a solid angle spread $\Delta
\Omega_S$. The four plots are for the four cases described in Fig. \ref{freqplots}.
All plots are made for the following source parameters:
 $\{\iota=\pi/4, \psi=\pi/3, \nu=3~{\rm mHz}\}$.
}\label{solidangle}
\end{figure}

\begin{figure}[!hbt]
\centerline{\psfig{file=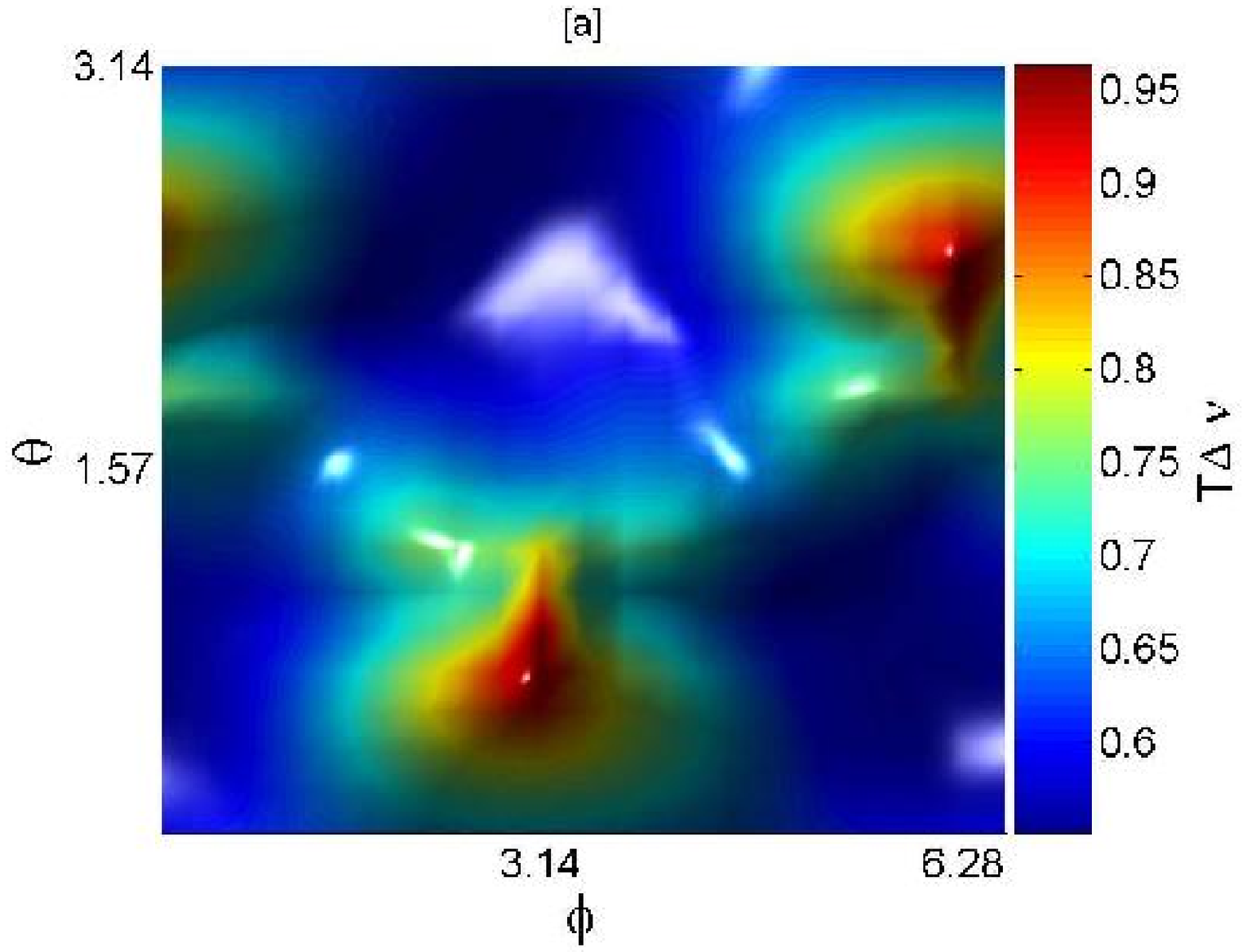,height=3.5in,width=3.5in}\psfig{file=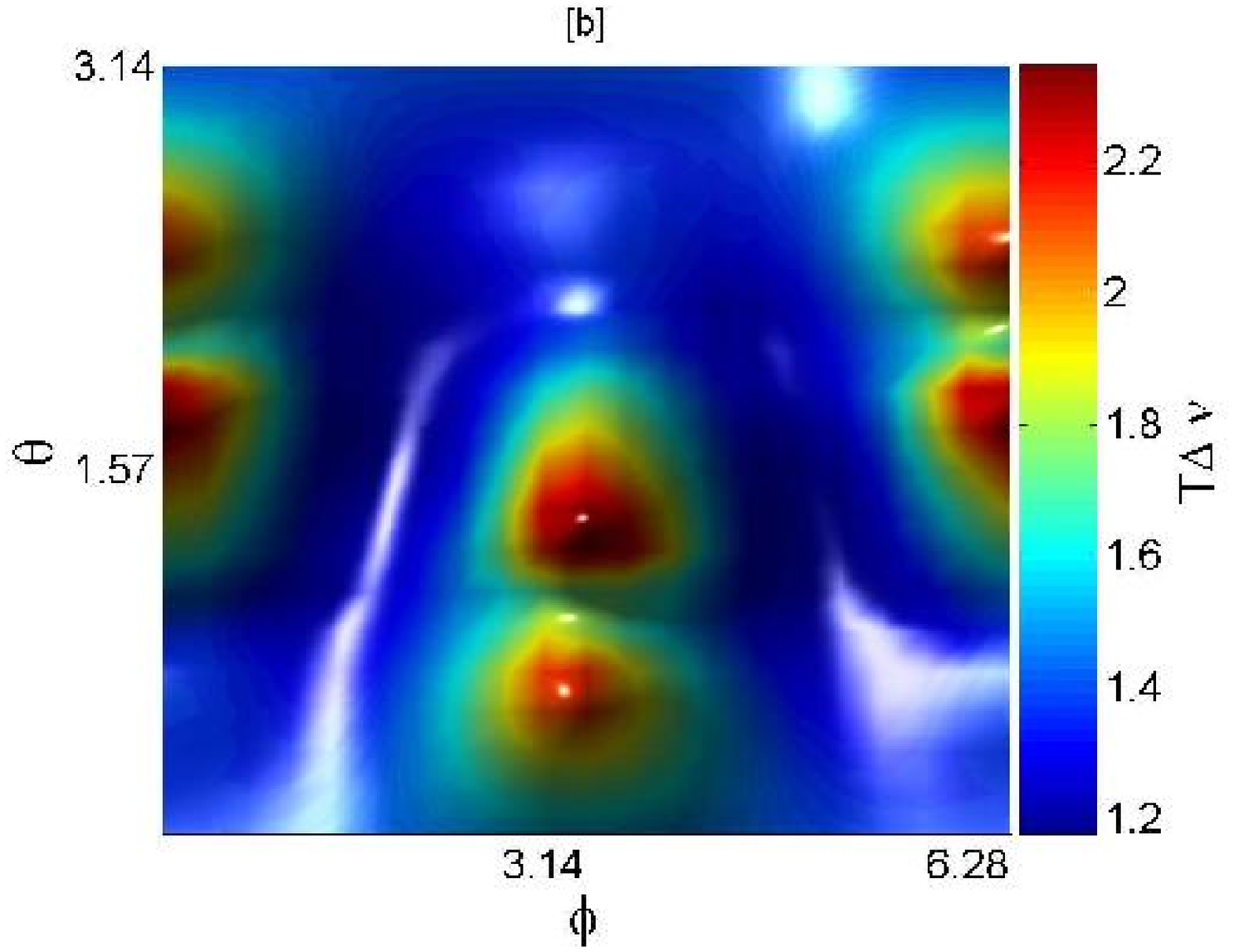,height=3.5in,width=3.5in}}
\centerline{\psfig{file=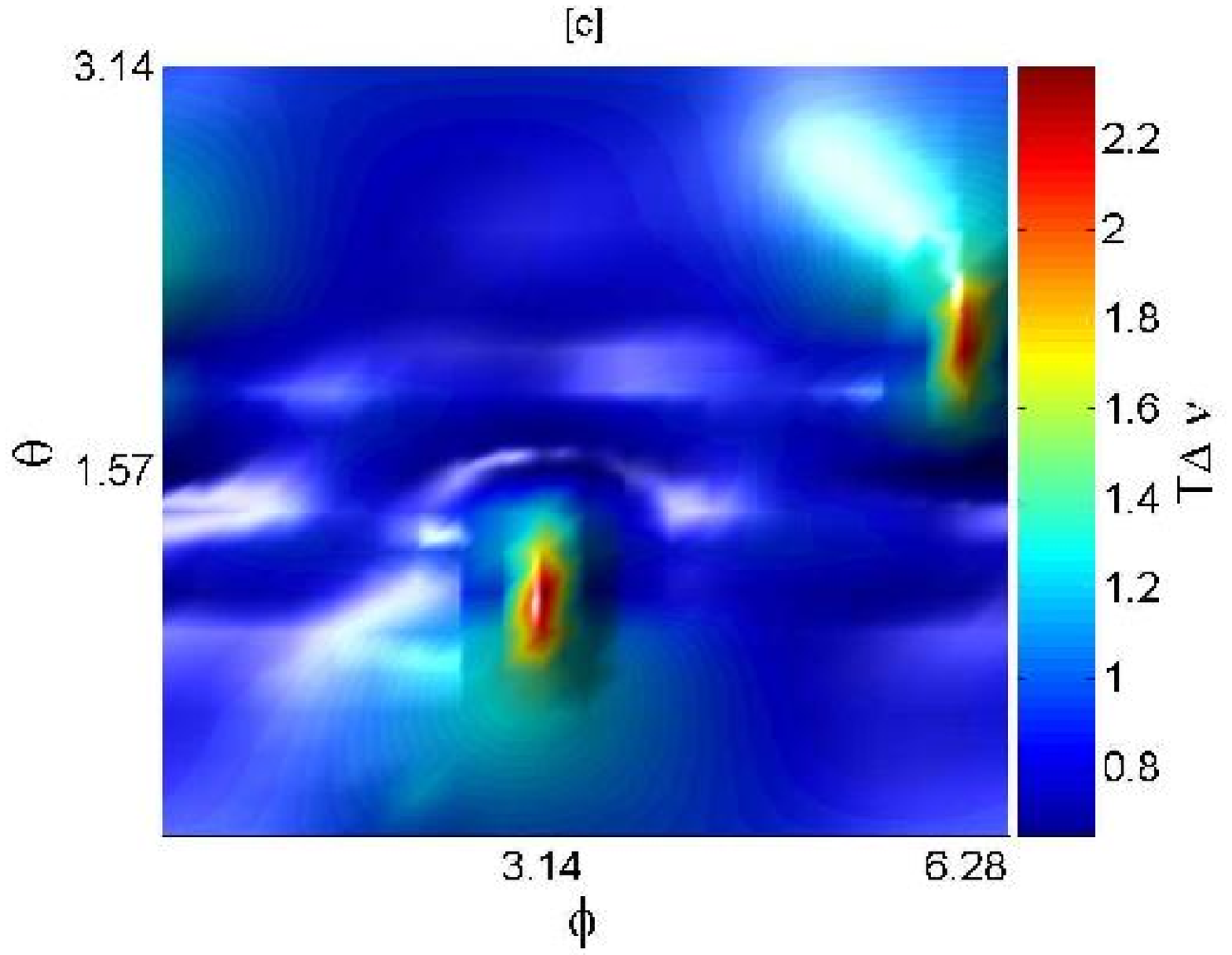,height=3.5in,width=3.5in}\psfig{file=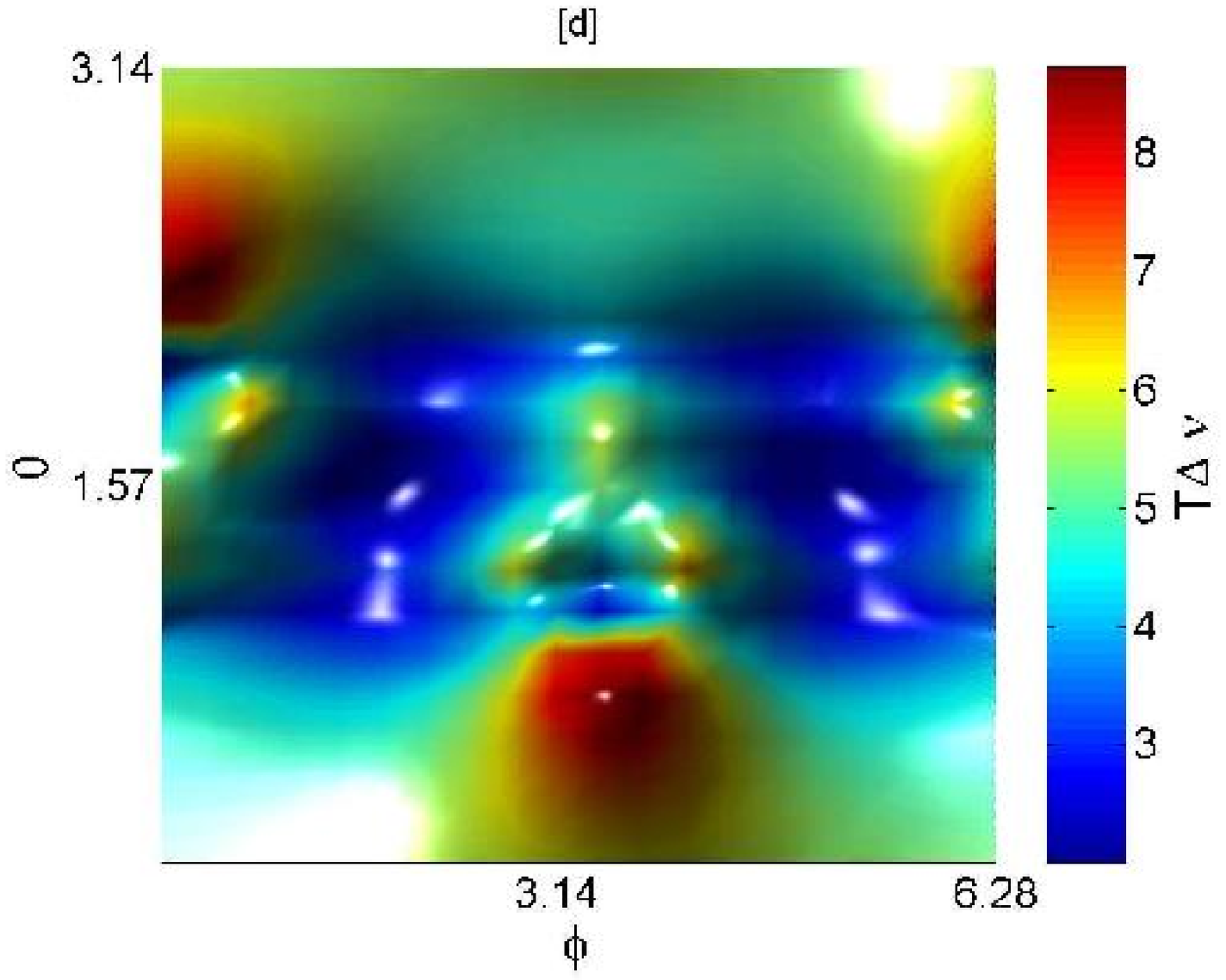,height=3.5in,width=3.5in}}
\caption{These are plots of the errors associated with determining
the dimensionless emission frequency parameter $T \Delta \nu$.
The four plots are for the four cases described in Fig. \ref{freqplots}.
All plots are made for the following source parameters:
 $\{\iota=\pi/4, \psi=\pi/3, \nu=3~{\rm mHz}\}$.}
\label{freqerror}
\end{figure}

\begin{figure}[!hbt]
\centerline{\psfig{file=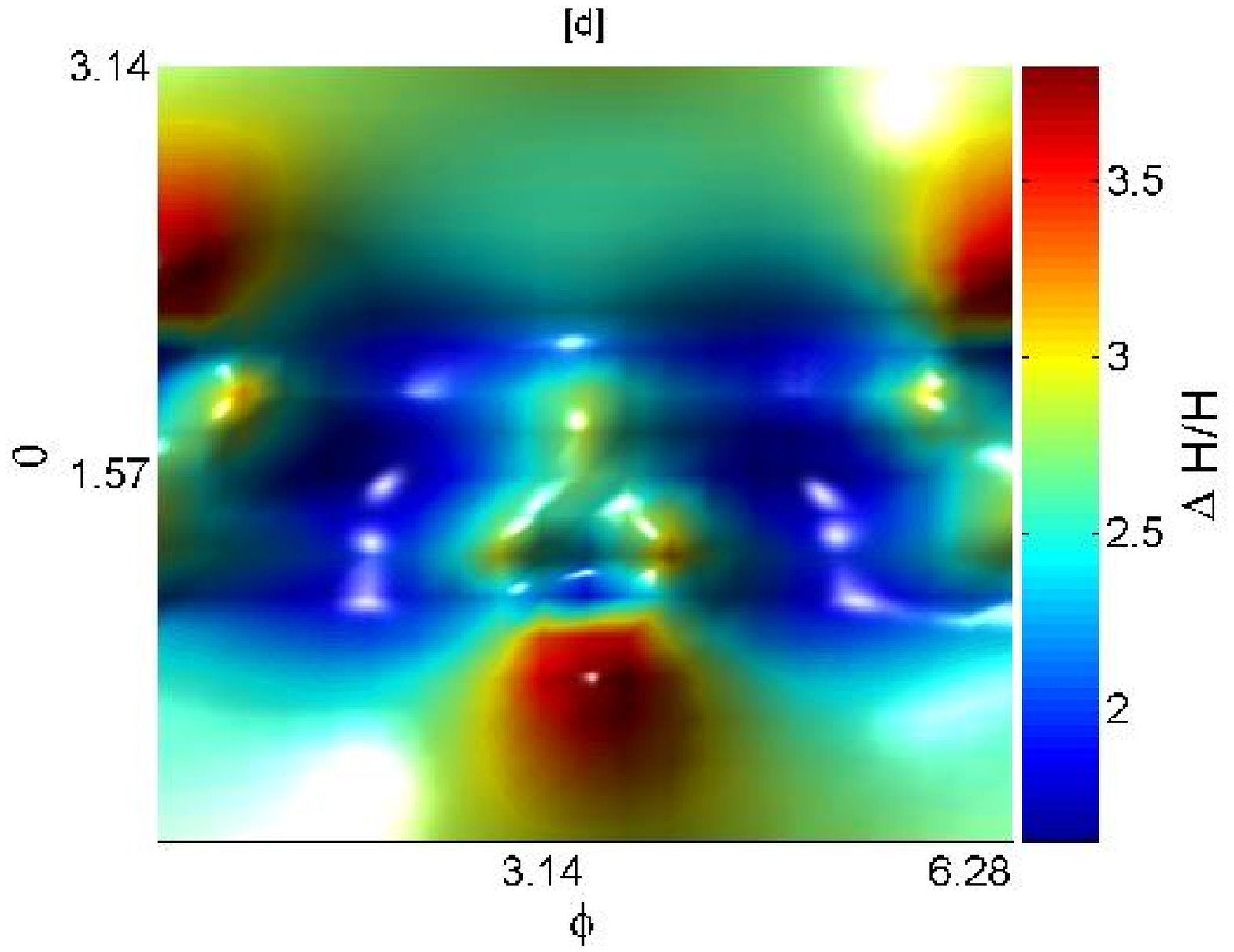,height=2.5in,width=2.5in}\psfig{file=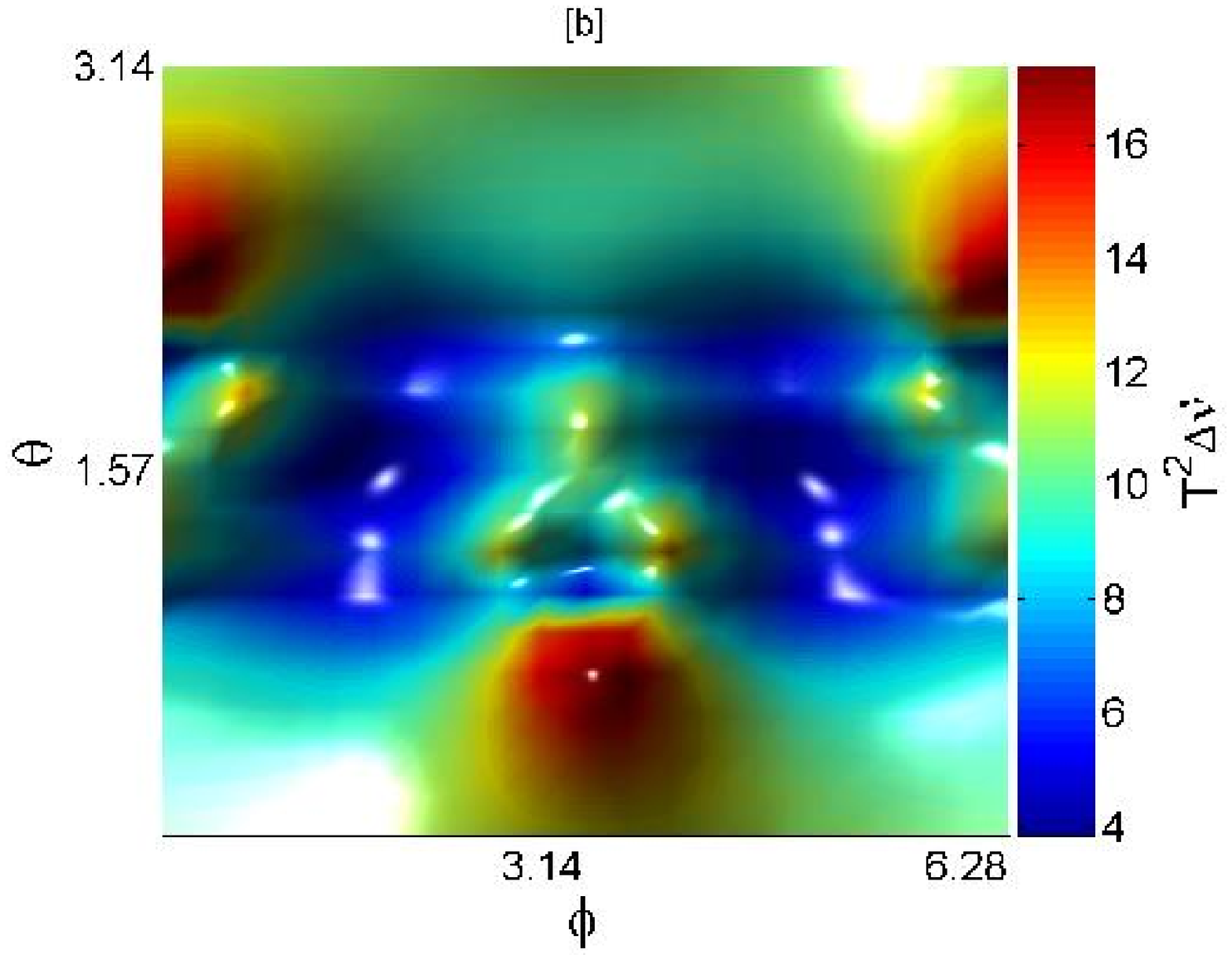,height=2.5in,width=2.5in}\psfig{file=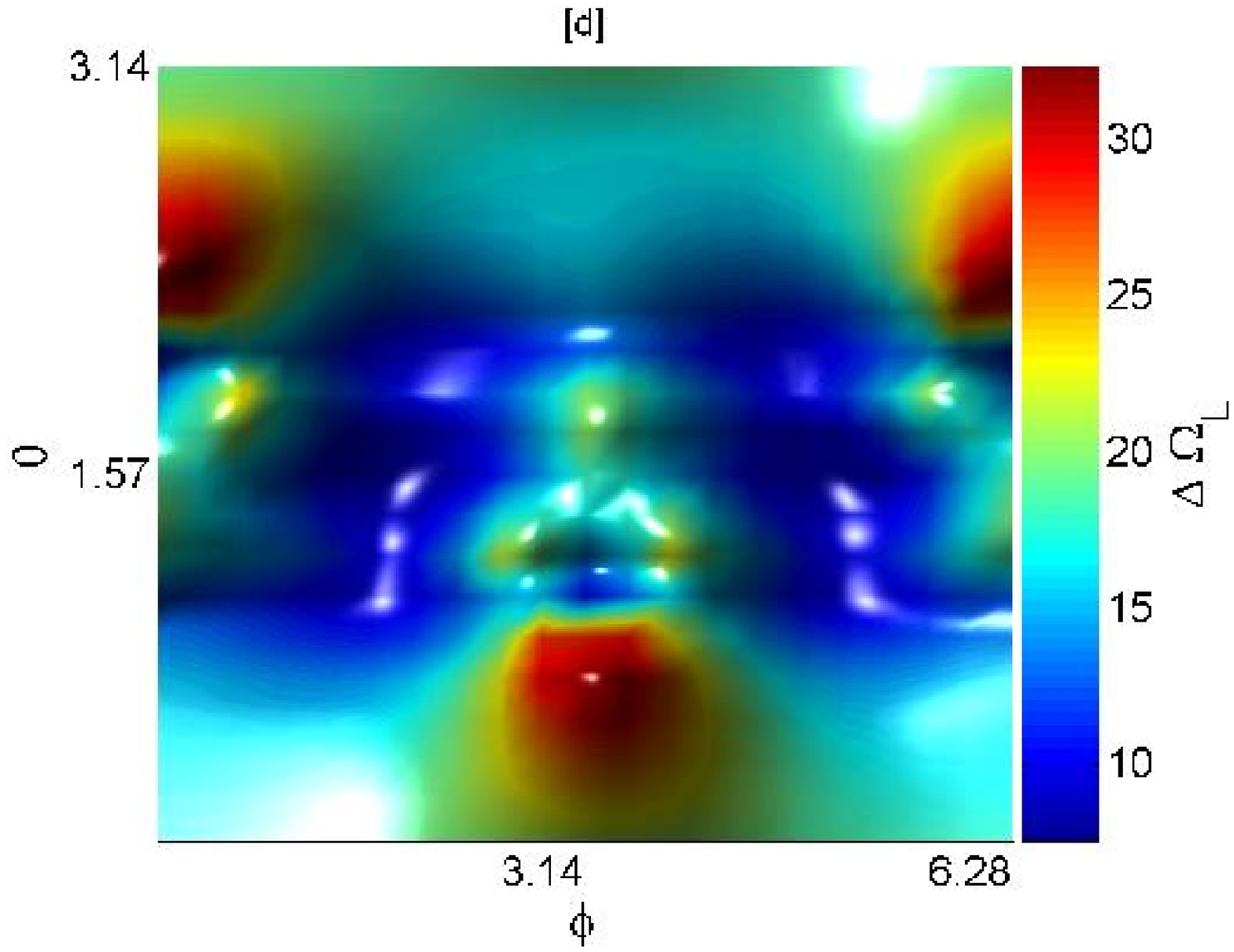,height=2.5in,width=2.5in}}
\caption{Plots of the errors associated with the remaining parameters: 
$\Delta H/H$, $T^2\Delta \dot{\nu}$, and $\Delta \Omega_L$. They all
correspond to case (d) described in Fig. \ref{freqplots}. Both Doppler-phase
modulation and frequency evolution are included here.
All plots are made for the following source parameters:
 $\{\iota=\pi/4, \psi=\pi/3, \nu=3~{\rm mHz}\}$.}
\label{amperror}
\end{figure}

We plot the parameter errors as functions of sky position in
Figs. \ref{solidangle}, \ref{freqerror}, and \ref{amperror} for the same 
source polarization state and frequency as described earlier in this
section. Note that these parameter-error
sky plots are presented for the $\bar{A}\bar{E}\bar{T}$ pseudo-detector
network. Altering $\nu$,
the polarization state, or the pseudo-detector network will, in general,
change the details found in these plots. However, certain qualitative aspects,
discussed below, will remain unaffected.

While all four cases (i.e., (a)-(d))
are plotted for sky-position and frequency errors in Figs. \ref{solidangle}
and \ref{freqerror}, respectively, only the full case (i.e., case (d)) is
plotted for the remaining parameters in Fig. \ref{amperror}. This is because
the plot for each case of these remaining parameters strongly resembles
the frequency-error sky plot of the corresponding case in Fig. \ref{freqerror}.
Indeed, a quick examination of all the error plots reveals that they can be
divided into two sets based on their apparent symmetry: (1) those for the
sky-position and (2) those for the rest of the parameters. The first set
displays a {\em strong} axisymmetry (i.e., symmetry with respect to translations
along the $\phi$ axis) that is missing in the other set.
The reason behind this divide is that
the Fisher information matrix and, therefore, the parameter covariance matrix,
is predominantly block diagonal, with
two blocks. The first block comprises just the sky-position ($\theta$, $\phi$)
variance-covariance and the second one comprises the variance-covariance of
the rest of the parameters. The elements outside of these two blocks are not all
zeros, but are much smaller than the elements within the blocks. Consequently,
any error in a parameter in one set influences that in another parameter from
the same set much more strongly than in the other set. This explains why the
patterns in each case
of the second set are so similar among all parameters within that set, as
is manifest in Fig. \ref{amperror}.
Also, the reason why the sky-position error plots are predominantly axisymmetric
while the other ones are not is because while the latter are affected by
the beam-pattern functions alone, the former are affected by them as well as
their parameter derivatives. It is worth noting that the deviation from axisymmetry
is still apparent in Figs. \ref{solidangle}, however, this effect is significantly reduced
with respect to the other parameters.

There is an alternative way of understanding the three ``unphysical'' cases
(i.e., cases (a)-(c)) studied in Figs. \ref{solidangle} and \ref{amperror}.
Starting with the last of these, case (c), note that
it is identical to the physical case where the chirping is
negligible. This is indeed the case when the chirp mass is as low as
0.3$M_\odot$. As seen in Fig. \ref{omegadotmismatch}, for such a case,
filtering the data
with chirp-less templates causes negligible SNR loss. Thus, the case (c)
plots in these figures indicate the expected random errors in such a search.

Case (b) resembles the physical case where the Doppler-phase
modulation is negligible, which occurs when the source frequency is below
a milli-Hertz (barring the slight ruining of the sky-position estimates).
Since the chirp rate was fixed in this plot, to compensate for
lowering the frequency to below 1 mHz one must increase the chirp mass,
which affects the signal amplitude but not the SNR-normalized error shown
here. Thus, case (b) is the expected error plot for all sources that have
frequencies less than about 1 mHz but chirp masses commensurately larger than
5.3$M_\odot$ (such that ${\cal M}_c^{5/3}\omega^{11/3}=$ constant).
Comparing the plots for cases (b) and (d) then shows that
the SNR-normalized error tends to increase at low frequencies, more so at
the poles than near the ecliptic.

Finally, case (a) represents the physical case where
$\nu < 1$ mHz and ${\cal M}_c$ is very small, such that
both Doppler-phase modulation and source chirping are negligible.
(Note that the sky-position
error is almost constant in the LISA band below 1 mHz (cf. Fig. \ref{freqplots}).) However as Fig.
 \ref{dopplermismatch} illustrates, this will not occur in the LISA band.

\section{\label{sec:conclusion}Conclusion}

In this paper we studied the effect of Doppler-phase modulation, frequency
evolution, and time-delay interferometry on the accuracy of estimating
signal parameters of low-mass circular binaries. All results presented here
were obtained for maximum-likelihood
estimates of binary signal parameters in a specific network of LISA's TDI
variables, namely, the $\bar{A}\bar{E}\bar{T}$ variables, and 
for a one-year integration time. We find that 
Doppler-phase modulation not only improves the sky-position estimation 
accuracy for $\nu \simgt 1$ mHz, but also degrades it in the range
$0.1 {\rm mHz} \simlt \nu \simlt 1$ mHz. It also increases the error in the
chirp parameter by a factor of 2 
and the frequency by a factor of 1.5 at $\nu \simeq 10$ mHz. It leaves the 
accuracies of the remaining parameters, namely, the source orientation and 
the signal amplitude mostly unaffected.

The exclusion of the chirp (or frequency evolution) parameter affects the
SNR loss by different amounts for binaries with different chirp masses.
For instance, Fig. \ref{omegadotmismatch} shows that for ${\cal M}_c \simlt 5.3~M_\odot$,
such an exclusion causes negligible effect on the SNR for $\nu \simlt 1$ mHz.
Since in this low-frequency band the estimation accuracy of all signal 
parameters, except $\nu$ and $\dot{\nu}$, is affected by its inclusion, 
from a few percent for the sky-position error to a factor of 6 for the 
source orientation, it is advisable
to drop the chirp parameter from the search templates for these sources.
On the other hand, since the SNR loss from excluding the chirp parameter
cannot be ignored when $\nu \simgt 3$ mHz,
for all but the lowest chirp masses of interest, one has to accept the
parameter-error consequences for these high frequency sources.
Luckily, a majority of the parameters remain unaffected in this high-frequency band, 
except $\nu$ and 
$\dot{\nu}$. 
Thus, if one were to turn off or on the inclusion of the chirp parameter
judiciously, so as to get the most accurate estimates 
possible for $\Omega_L$, $\Omega_S$, $H$, then for a given chirp mass, the error profiles in the
frequency plots would be such that they are low (i.e., at the levels shown 
in Fig. \ref{omegadotmismatch}a) at the low frequencies, but rise at 
the mid-frequencies, and 
fall off at the high frequencies as in Fig. \ref{omegadotmismatch}d.
The biggest casualty arising from searching for the chirp at high 
frequencies is $\dot{\nu}$: its error can suffer
by as much as a factor of 5 at frequencies greater than 3 mHz (see
Fig. \ref{freqplots}d). A similar conclusion was arrived at earlier 
by Takahashi
and Seto \cite{Takahashi:02} while working with the Michelson variables,
although their exact numbers were somewhat different. Clearly, this will
have an adverse effect on the resolvability of the confusion noise
above 3 mHz.

Comparing the frequency plots we obtained here to those found in 
Ref. \cite{Takahashi:02}, one striking aspect is the reduced levels for 
the SNR-normalized errors we find for most parameters at all frequencies.
Its origin lies in the two principal differences between our studies, namely, the choice of the data variables (TDI versus Michelson variables) and the
approximations made in Ref. \cite{Takahashi:02}.
For instance, the value of $\Delta H/H$ presented in \cite{Takahashi:02}
is its average over LISA's band, whereas we derive the values of 
$\Delta H/H$ for different frequencies. Also, we do not use the long-wavelength
approximation, which has been shown to have significant effects at frequencies
above 3mHz \cite{Vecchio:2004ec}.

One of the many interesting astrophysical sources that LISA can target 
is a stellar-mass binary containing a black hole. It is unclear how many 
such objects are present in our galaxy.
These objects will likely have chirp masses in excess of 2$M_\odot$. Figure
\ref{omegadotmismatch} shows that for $\nu \simgt 2$ mHz, detecting
them will require inclusion of the chirp parameter in the search 
templates. This, however, will
result in large inaccuracies in all of its parameters in the range 2-3 mHz.
That in turn will hamper our ability to discern their signals from those of
the confusion noise, unless their SNRs are sufficiently large. A large SNR
can result either from the proximity of a source or a large chirp mass or both.
A large chirp mass, however, will not always serve this purpose 
because the 2-3 mHz 
range actually widens towards lower frequencies with increase in chirp mass.
This is because the point at which $m(\mbox{\boldmath $\vt$},\mbox{\boldmath $\vt$}')$ starts deviating from unity in Fig. \ref{omegadotmismatch}
shifts to lower frequencies with increasing ${\cal M}_c$. This identifies 
a problem with LISA's ability to detect stellar mass black hole binaries,
about which very little is known in terms of their formation and demographics.

Note that the above conclusions will change, in general, as the integration
time is increased. For example, a ten-year integration time will not only
improve the accuracy with which the chirp parameter will be determinable,
but also reduce the adverse effects that its inclusion in search templates
has on the estimation accuracy of other parameters \cite{Takahashi:02}.
Some changes may also appear in the SNR-normalized frequency and 
sky plots if one were to make them for a different TDI network. However,
the major change in that context will arise from the different
sensitivities that the different TDI networks have to different 
sky-positions, thus, affecting the absolute (unnormalized) error values. 
These aspects will be explored in more details elsewhere \cite{BoseRoganPrep1}.

\acknowledgments

We would like to thank Curt Cutler for helpful discussions. Thanks are also
due to Michael Stoops, Svend Sorensen, Nathan Hearn and George Lake for help with
computational resources. This work was funded in part by NASA Grant NAG5-12837.


\begin{thebibliography}{9}

\bibitem{PrePhaseA} P.~L. Bender, et al., \textit{LISA Pre-Phase A Report; Second Edition}, MPQ 233 (1998).

\bibitem{Abbott:2003vs}
  B.~Abbott {\it et al.}  [LIGO Scientific Collaboration],
  Nucl.\ Instrum.\ Meth.\ A {\bf 517}, 154 (2004)
  [arXiv:gr-qc/0308043].

\bibitem{VIRGO}
F.~Acernese {\it et al.}, Class.\ Quant.\ Grav.\  {\bf 22}, S869 (2005).

\bibitem{TAMA}
M.~Ando {\it et al.} [TAMA Collaboration],  Class.\ Quant.\ Grav.\  {\bf 22}, S881 (2005).

\bibitem{Willke:2002bs}
  B.~Willke {\it et al.},
  Class.\ Quant.\ Grav.\  {\bf 19}, 1377 (2002).

\bibitem{Mio:2003ii}
 N.~Mio  [LCGT Collaboration],
  Prog.\ Theor.\ Phys.\ Suppl.\  {\bf 151}, 221 (2003).

\bibitem{Cutler:2002me}
  C.~Cutler and K.~S.~Thorne,
  arXiv:gr-qc/0204090.

\bibitem{Nayak:2003na}
K.~R.~Nayak, S.~V.~Dhurandhar, A.~Pai and J.~Y.~Vinet,
Phys.\ Rev.\ D {\bf 68}, 122001 (2003) [arXiv:gr-qc/0306050].

\bibitem{Bender:1997hs}
P.~L.~Bender and D.~Hils,
Class.\ Quant.\ Grav.\  {\bf 14}, 1439 (1997).

\bibitem{hils} D. Hils \& P. L. Bender, ApJ {\bf 537}, 334 (2000).

\bibitem{nelemans} G. Nelemans, L. R. Yungelson \& S. F. Portegies Zwart, A\&A {\bf 375}, 890 (2001).

\bibitem{gclean} N.J. Cornish \& S.L. Larson, Phys. Rev. D{\bf 67}, 103001 (2003).

\bibitem{mohanty} M. S. Mohanty, \& R. K. Nayak, gr-qc/0512014 (2005).

\bibitem{Crowder:2006wh}
  J.~Crowder, N.~J.~Cornish and L.~Reddinger,
  Phys.\ Rev.\ D {\bf 73}, 063011 (2006)
  [arXiv:gr-qc/0601036].

\bibitem{Cutler:98}
C.~Cutler, Phys.\ Rev.\ D {\bf 57}, 7089 (1998).

\bibitem{Takahashi:02}
R.~Takahashi and N.~Seto, Astrophy. J. {\bf 575}, 1030 (2002).

\bibitem{Schutz}
B.~F.Schutz, Nature {\bf 323}, 310 (1986).

\bibitem{Barack:2003fp}
  L.~Barack and C.~Cutler,
  Phys.\ Rev.\ D {\bf 69}, 082005 (2004)
  [arXiv:gr-qc/0310125].

\bibitem{Vecchio:2004ec}
  A.~Vecchio and E.~D.~L.~Wickham,
  Phys.\ Rev.\ D {\bf 70}, 082002 (2004)
  [arXiv:gr-qc/0406039].

\bibitem{Tinto:yr}
M.~Tinto and J.~W.~Armstrong,
Phys.\ Rev.\ D {\bf 59}, 102003 (1999).

\bibitem{Armstrong:99}
J.~W.~Armstrong, F.~B.~Estabrook, and M.~Tinto, Astrophys. J., {\bf
527}, 814 (1999).

\bibitem{TEA}
M.~Tinto, F.~B.~Estabrook, and J.~W.~Armstrong,  
Phys.\ Rev.\ D {\bf 65}, 082003 (2002).

\bibitem{Hels}
C.~W.~Helstrom, {\sl Statistical Theory of Signal Detection}
(Pergamon Press, London, 1968).

\bibitem{lisaSite}
High resolution versions of figures presented in this paper are available
at http://gravity.physics.wsu.edu/LISA/parameterEstimates/.

\bibitem{indexing}
Note that the
indices $i$ and $i\pm 1$ can take only 1, 2, and 3 as values. These
three numbers are ordered clockwise in Fig. \ref{LISAtri}. By
convention, whereas $i+1$ equals the number next to $i$ while going
clockwise in that figure, $i-1$ equals the number preceding $i$.
E.g., when $i=3$, we take $i-1=2$ and $i+1=1$; when $i=1$, we take
$i-1=3$ and $i+1=2$.

\bibitem{Rogan:2004wq}
  A.~Rogan and S.~Bose,
  Class.\ Quant.\ Grav.\  {\bf 21}, S1607 (2004)
  [arXiv:gr-qc/0407008].

\bibitem{Dhurandhar:2001kx}
S.~V.~Dhurandhar, K.~Rajesh Nayak and J.~Y.~Vinet,
Phys.\ Rev.\ D {\bf 65}, 102002 (2002) [arXiv:gr-qc/0112059].

\bibitem{Prince:2002hp}
~A.~Prince, M.~Tinto, ~L.~Larson and ~W.~Armstrong,
Phys.\ Rev.\ D {\bf 66}, 122002 (2002) [arXiv:gr-qc/0209039].

\bibitem{DhurandharCombo}
These combinations are identical to  $X^{(3)}$, $X^{(4)}$, and 
$-X^{(1)}+\zeta_3X^{(2)}$, respectively, used by Dhurandhar et al. in 
Ref. \cite{Dhurandhar:2001kx}.

\bibitem{RoganCorr1}
There is a typographical error in Eq. (18) of Ref. \cite{Rogan:2004wq}:
In the expression for $P^{(3)}(f)$, the $\cos(2\pi fL)$ factor should appear 
as squared.

\bibitem{Cornish:2003tz}
N.~J.~Cornish and R.~W.~Hellings,
Class.\ Quant.\ Grav.\  {\bf 20}, 4851 (2003) [arXiv:gr-qc/0306096].

\bibitem{Shaddock:2003dj}
D.~A.~Shaddock, M.~Tinto, F.~B.~Estabrook and J.~W.~Armstrong,
Phys.\ Rev.\ D {\bf 68}, 061303 (2003) [arXiv:gr-qc/0307080].

\bibitem{Tinto:2003vj}
M.~Tinto, F.~B.~Estabrook and J.~W.~Armstrong,
Phys.\ Rev.\ D {\bf 69}, 082001 (2004) [arXiv:gr-qc/0310017].

\bibitem{Tinto:2003gwdaw}
M.~Tinto,
8th Anual Gravitational Wave Data Analysis Workshop, (2003).

\bibitem{MTW}
C.~W.~Misner, K.~Thorne, J.~A.~Wheeler, {\sl Gravitation} (W. H. Freeman, New York, 1973)).

\bibitem{GMS}
M.~Gel'fand, R.~A.~Minlos, and Z.~Ye.~Shapiro, {\sl Representations of the 
Rotation and Lorentz Groups and their Applications} (Pergamon Press, New York,
1963).

\bibitem{Bose:1999pj}
S.~Bose, A.~Pai and S.~V.~Dhurandhar,
Int.\ J.\ Mod.\ Phys.\ D {\bf 9}, 325 (2000) [arXiv:gr-qc/0002010].

\bibitem{Pai:2000zt}
A.~Pai, S.~Dhurandhar and S.~Bose,
Phys.\ Rev.\ D {\bf 64}, 042004 (2001) [arXiv:gr-qc/0009078].

\bibitem{KrolakRef}
The same problem was addressed independently by using a different method in
Ref. \cite{Krolak:2004xp}.

\bibitem{Krolak:2004xp}
A.~Krolak, M.~Tinto and M.~Vallisneri, ``Optimal filtering of the
LISA data,'' [arXiv:gr-qc/0401108].

\bibitem{Jaranowski:2005hz}
  P.~Jaranowski and A.~Krolak,
  Living Rev.\ Rel.\  {\bf 8}, 3 (2005).

\bibitem{Seto:2001pg}
N.~Seto,
Phys.\ Rev.\ Lett.\  {\bf 87}, 251101 (2001)
[arXiv:astro-ph/0111107].

\bibitem{BoseRoganPrep1}
S.~Bose and A.~Rogan, {\it In preparation}.

\end{thebibliography}
\end{document}